\documentclass{article}

\usepackage{jcappub} 
\usepackage[english]{babel}
\usepackage[letterpaper,top=2cm,bottom=2cm,left=3cm,right=3cm,marginparwidth=1.75cm]{geometry}

\usepackage{xcolor}
\usepackage{bm}
\usepackage{ulem}
\usepackage{graphicx}
\usepackage{geometry}
\usepackage{multirow}
\usepackage{amsmath}
\usepackage{mathtools}
\usepackage{blindtext}

\usepackage{babel}
\usepackage{soul}

\title{\boldmath{
Lowering the strong coupling mode of modified teleparallel gravity theories
}}

\author[a,b]{Yu-Min Hu, }
\author[a,b]{Bi-Chu Li, }
\author[c]{Yang Yu, }
\author[d,a,1]{Martin Kr\v{s}\v{s}\'ak, }
\author[e,a,f,1]{Emmanuel N. Saridakis}
\author[a,b,1]{Yi-Fu Cai \note{To whom correspondence should be addressed.}}

\affiliation[a]{Department of Astronomy, School of Physical Sciences, University of Science and Technology of China, 96 Jinzhai Road, Hefei, Anhui 230026, China}
\affiliation[b]{CAS Key Laboratory for Research in Galaxies and Cosmology, School of Astronomy and Space Science,
University of Science and Technology of China, 96 Jinzhai Road, Hefei, Anhui 230026, China}
\affiliation[c]{School of Physics and Astronomy, Sun Yat-sen University, Zhuhai 519082, China}
\affiliation[d]{Department of Theoretical Physics, Faculty of Mathematics, Physics and
Informatics, Comenius University in Bratislava, 84248, Slovak Republic}
\affiliation[e]{National Observatory of Athens, Lofos Nymfon, 11852 Athens, 
Greece}
\affiliation[f]{Departamento de Matem\'{a}ticas, Universidad Cat\'{o}lica del 
Norte,  Avda. Angamos 0610, Casilla 1280 Antofagasta, Chile}

\emailAdd{yumin28@ustc.edu.cn}
\emailAdd{yuyang69@mail2.sysu.edu.cn}
\emailAdd{martin.krssak@gmail.com}
\emailAdd{msaridak@noa.gr}
\emailAdd{yifucai@ustc.edu.cn}

\abstract{ 
We investigate the strong coupling problem in modified teleparallel gravity 
theories using the effective field theory (EFT) approach, demonstrating that it 
is possible to shift the emergence of new degrees of freedom (DoFs) to lower orders in 
perturbation theory. We first focus on the case of $f(T)$ gravity, and  we 
show that in its 
conformally equivalent form  the scalar perturbations are 
non-dynamical up to the cubic action.
We then propose a simple modification of the theory, which lowers  the 
appearance of new DoFs to cubic order, compared to the quartic 
order in standard $f(T)$ gravity. Our work  opens a new avenue  to address 
the issue of strong coupling in 
modified teleparallel gravity,  and suggests a new
classification scheme  of these theories based on the perturbative order at 
which new DoFs appear. 
}
\begin{document}
\maketitle

\section{Introduction}

Modified gravity theories have attracted  significant attention as a possible 
solution to explain the accelerated expansion of the Universe and understand the 
physics beyond the standard cosmological model 
\cite{CANTATA:2021ktz,Carroll:2003wy,Nojiri:2003ft,Copeland:2006wr}.   
A particularly  interesting case  is the  class of modified teleparallel 
theories, where we modify not the standard general relativity but its 
teleparallel formulation, where gravity is described using teleparallel geometry 
and is attributed to the connection torsion 
\cite{Aldrovandi:2013wha,Cai:2015emx,Krssak:2018ywd,Bahamonde:2021gfp}. The most 
popular among these theories is the   $f(T)$ gravity one
\cite{Ferraro:2006jd,Ferraro:2008ey,Bengochea:2008gz,Linder:2010py}, where the 
Lagrangian is taken to be a non-linear function of the torsion scalar $T$ 
defining the teleparallel equivalent of general relativity (TEGR) 
\cite{Aldrovandi:2013wha,Maluf:2013gaa}. Since the torsion scalar contains only 
first derivatives of the dynamical variables, $f(T)$ gravity automatically 
avoids the problem of Ostrogradsky instabilities encountered in some 
curvature-based theories of gravity. The torsion scalar differs from the 
Levi-Civita Ricci scalar by a total derivative, and hence   $f(T)$ gravity is 
generally not equivalent to $f(R)$ gravity or any other curvature-based theory 
of gravity. This class of theories was extensively studied \cite{Zheng:2010am, 
Cai:2011tc, Cardone:2012xq, Krssak:2015oua, Chen:2019ftv, Ren:2021uqb, 
Ren:2021tfi, dosSantos:2021owt, Zhao:2022gxl, Wang:2023qfm}, and many other 
modified teleparallel models were proposed in recent years 
\cite{Geng:2011aj,Geng:2011ka,Otalora:2014aoa,Bahamonde:2016kba,
Bahamonde:2017wwk,Hohmann:2017duq,Hohmann:2018dqh,Hohmann:2018rwf,
Bahamonde:2022ohm}. Moreover, an entirely different branch of modified 
teleparallel theories has appeared based on the symmetric teleparallelism, where 
gravity is attributed to the non-metricity  
\cite{Nester:1998mp,Adak:2005cd,BeltranJimenez:2017tkd,BeltranJimenez:2019tme,
Gakis:2019rdd,Lazkoz:2019sjl,Zhao:2021zab,Heisenberg:2023lru,Boehmer:2023knj}.

While $f(T)$ gravity  may naively appear to be a simple theory,  its dynamics 
is extraordinarily  difficult to understand due to its highly non-linear 
character \cite{Li:2011rn, Ferraro:2018axk, Blixt:2020ekl, Blagojevic:2020dyq}, 
and the issue of disappearance of propagating modes at the linear level  in 
highly symmetric spacetimes  \cite{Dent:2010nbw,Chen:2010va, 
Izumi:2012qj,Chen:2014qtl,BeltranJimenez:2020lee,BeltranJimenez:2021auj, 
Shabani:2023xfn, Gomes:2023tur, Golovnev:2018wbh, Hohmann:2020vcv, 
Golovnev:2020aon}. This discontinuity in the number of modes 
is commonly referred to as the strong coupling problem and is encountered in 
other gravity models, where it is generally considered problematic and 
associated with the appearance of ghost-like modes \cite{Deffayet:2005ys, 
Blas:2009yd,Cai:2009dx}. Let us remark here that  the problem of strong coupling 
appears in the symmetric teleparallel modified theories of gravity, such as 
$f(Q)$ gravity model, as well \cite{Hu:2022anq, Tomonari:2023wcs, 
Hu:2023gui,Zhao:2024kri}.

From the effective  field theory (EFT) perspective, the  question is at what 
energy scales this strong coupling issue arises and whether it can be shifted 
beyond a certain cutoff. Using the torsional EFT approach \cite{Li:2018ixg, 
Cai:2018rzd, Cai:2019bdh, Yan:2019gbw, Ren:2022aeo, Mylova:2022ljr, Hu:2023xcf, 
Yang:2024kdo}, an energy scale analysis based on comparisons with linear and 
second-order scalar perturbations was provided and suggested that such strong 
coupling in $f(T)$ gravity may only exhibit strong nonlinear effects at 
extremely high energy scales for subhorizon modes \cite{Hu:2023juh}. Therefore, 
this kind of strong coupling issue may not lead to truly catastrophic 
consequences from an EFT point of view.

In this work  we will extend this previous analysis \cite{Hu:2023juh} to the 
conformal equivalent form of the $f(T)$ gravity and the new models derived from 
it.  A conformal transformation is a transformation  where the fields are 
rescaled by some conformal factor $\Omega(x)$, which leaves the light cones and 
hence the causal structure unchanged \cite{dicke1962mach,Maeda:1988ab, 
Deruelle:2010ht, Paliathanasis:2023gfq, 
Yang:2010ji, Hohmann:2018vle, Bamba:2013jqa}.  The reason for considering 
conformal transformations is that two systems connected by an invertible transformation (excluding potential singular cases \cite{Jirousek:2022jhh}) are 
dynamically equivalent, although the physical nature of this equivalence  is 
a matter of ongoing discussion 
\cite{Flanagan:2004bz,Faraoni:2006fx,Chiba:2013mha,Kamenshchik:2014waa}. By choosing 
particular conformal transformations associated with field-dependent rescaling, 
one can effectively map one class of theories into a more convenient form for 
study, particularly in scenarios involving kinetic mixing between the 
metric/tetrads and other fields. The well-known example is the case of $f(R)$ 
gravity, which is a non-minimally coupled theory in the Jordan frame, and using 
conformal transformations can be brought into minimally coupled form in the 
Einstein frame \cite{Faraoni:1998qx,DeFelice:2010aj}. In the case of $f(T)$ 
gravity, the non-minimal coupling between torsion as well as the auxiliary 
scalar field cannot be completely removed  using a conformal transformation, 
nevertheless it allows us to bring it to the form where the scalar field 
couples only to the vectorial torsion \cite{Wright:2016ayu,Jarv:2018bgs}, 
which we call here the conformal equivalent of $f(T)$ gravity.

We show here  that a conformal equivalent of $f(T)$ gravity could serve as a 
useful tool for dealing with nonlinear perturbations, as the one-to-one mapping 
under conformal transformations does not change the number of DoFs 
\cite{Zumalacarregui:2013pma, Domenech:2015tca, Takahashi:2017zgr}. We first 
examine whether the behavior of scalar perturbations in $f(T)$ theory, which 
indicates no propagating modes up to the second order, is consistently reflected 
in one of its conformal equivalents using the EFT method. As a result, the 
underlying strong coupling behavior of scalar modes is similarly manifested in 
both conformal frames, at least at this given order. Additionally, we discuss 
the impact of the coupling term, naturally introduced by conformal 
transformations, on the strong coupling behavior by artificially modifying the 
coefficient.

This article is organized as follows.  In Sec. \ref{Section:2}, we briefly 
introduce the $f(T)$ theory and one of its conformal forms, as well as explain 
how they are incorporated into the framework of torsional EFT to facilitate 
subsequent analysis and calculations.
In Sec. \ref{Section:3}, we study the cosmological scalar  perturbations of 
$f(T)$ conformal equivalent theory around a flat 
Friedmann-Lema\^itre-Robertson-Walker (FLRW) background and investigate whether 
the strong coupling issue has been shown up to non-linear order. In Sec. 
\ref{Section:4}, we modify the coupling term in $f(T)$ conformal equivalent 
theory. 
Our results are summarized in Sec. \ref{Section:5}. 
Throughout the paper, we use unit $\frac{1}{8\pi G}=M^2_P$ with  the Planck mass 
$M_P$ and convention for the metric $\{-,+,+,+\}$. The operators 
$\overset{\circ}{\nabla}_\mu$ and $\nabla$ represent the covariant derivative 
adapted to the Levi-Civita connection and the general linear connection, 
respectively.

\section{\texorpdfstring{$f(T)$}{f(T)} gravity and its conformal equivalent 
theory}\label{Section:2}

In this section, we review teleparallel geometry and modified gravity theories 
with a   focus on  $f(T)$ gravity 
\cite{Aldrovandi:2013wha,Krssak:2018ywd,Krssak:2024xeh}. Additionally, we 
introduce the conformal equivalent formalism of $f(T)$ theory and show how it 
can be translated into EFT form. 

\subsection{Teleparallel geometry and \texorpdfstring{$f(T)$}{f(T)} gravity}

Teleparallel theories are  formulated within the tetrad formalism where at each 
point of the manifold we choose an orthonormal basis  $e_A$ for the tangent 
space called the tetrad,  and the corresponding co-basis $e^A$ called the 
cotetrad. We can then find the components of tetrads in the coordinate basis as 
$e_{A}= e^{\phantom{A}\mu}_A\partial_{\mu}$,  and analogously $e^{A}= 
e^A_{\phantom{A}\mu} {\rm d}x^{\mu}$ for the cotetrads. The covariant 
differentiation can be defined for both the spacetime and tangent space tensors. 
For a vector $X^A=e^A{}_\mu X^\mu$, we define the covariant derivative as  
\begin{equation}
	{\nabla}_\mu X^\rho=\partial_\mu X^\rho + {\Gamma}^\rho{}_{\nu\mu}X^\nu~, \qquad
	{\nabla}_\mu X^A=\partial_\mu X^A + {\omega}^A{}_{B\mu}X^B~,\label{covdef}
\end{equation} 
where $\Gamma^\rho{}_{\nu\mu}$ are the linear connection coefficients of the  
covariant derivative in the coordinate basis, and $\omega^A{}_{B\mu}$ are the 
spin connection coefficients of the covariant derivative in the non-coordinate 
(tetrad) basis. Both connection coefficients are related through
\begin{align}\label{tetpost}
 \partial_{\mu} e_{\ \nu}^{A} -\Gamma_{\ \nu\mu}^{\lambda} e_{\ \lambda}^{A} +\omega_{\ B\mu}^{A} e_{\ \nu}^{B} = 0 ~.
\end{align}
The linear connection coefficients do transform in a non-tensorial way under  
the change of coordinate basis, while under local Lorentz transformations of the 
cotetrad 
\begin{equation}\label{key}
 e^A{}_\mu = \Lambda^A{}_B e^B{}_\mu~, 
\end{equation}
the spin connection transforms as
\begin{equation}\label{spintransf}
 \omega^{A}_{\ B\mu} \rightarrow \Lambda^{A}_{\ C}(\Lambda^{-1})^{D}_{\ B}\omega^{C}_{\ D\mu} +\Lambda^{A}_{\ C}\partial_{\mu}(\Lambda^{-1})^{C}_{\ B}~.
\end{equation}
We can define various connections on the manifold, which  will lead generally  
to  metric-affine geometries, where the connection is characterized by the 
curvature, torsion and non-metricity tensors \cite{Hehl:1994ue,Obukhov:2002tm}. 
We then consider two special cases.

The first case is that of Riemannian or Levi-Civita geometry,  where the 
covariant derivative $\overset{\circ}{\nabla}_\mu$ is defined by vanishing 
torsion and non-metricity. This  uniquely determines the Levi-Civita linear 
connection coefficients to be given by Christoffel symbols 
$\overset{\circ}{\Gamma}{}^\lambda{}_{\mu\nu}$, from where we can derive the 
Levi-Civita spin connection using \eqref{tetpost} and calculate the Riemannian 
curvature $\overset{\circ}{R}{}^\rho{}_{\sigma\mu\nu}$. The corresponding scalar 
curvature 
$\overset{\circ}{R}{}=\overset{\circ}{R}{}^\rho{}_{\sigma\rho\nu}g^{\sigma\nu}$ 
is then used to define the Einstein-Hilbert Lagrangian of  general relativity, 
as well as other curvature-based modified theories of gravity, such as $f(R)$ 
gravity.

The second geometry of interest is teleparallel geometry,  where the connection 
is characterized by vanishing curvature and non-metricity, which results in the 
spin connection being given by the pure gauge connection\footnote{From now on, 
the bare geometric quantities will represent geometric objects from teleparallel 
geometry, e.g. $\omega^A_{\ B\mu}$ will be a teleparallel spin connection.} 
\begin{equation}
 \omega^A_{\ B\mu}=\Lambda^A_{\ C} \partial_\mu (\Lambda^{-1})^C{}_B ~.
\label{telcon}
\end{equation}
The reason why this is called  the pure gauge connection is that we can generate 
it from a zero connection using the transformation properties of the spin 
connection \eqref{spintransf}. If to each tetrad corresponds some pure gauge 
connection \eqref{telcon}, we can always consider an inverse local Lorentz 
transformation and transform the teleparallel spin connection to zero 
$\omega^{A}{}_{B\mu}\equiv 0$, known as the Weitzenb\"ock gauge 
\cite{Obukhov:2002tm}, which will now hold only is a special class of proper or 
good tetrads \cite{Ferraro:2011us,Tamanini:2012hg,Krssak:2015oua}. While the 
spin connection does indeed play a role in teleparallel theories, where it 
regularizes the action and covariantizes the theory, the general consensus is 
that we are allowed to set the Weitzenb\"ock gauge,  as long as we are 
interested in the dynamics of these theories and providing that we can guarantee 
to work in the class of proper or good tetrads.

The teleparallel linear connection in the Weitzenb\"ock gauge can  be obtained 
from \eqref{covdef} and is given by
\begin{align}
 \Gamma^\lambda{ }_{\nu \mu}=e_A{}^\lambda \partial_\mu e^A{ }_\nu~.
\end{align}
The curvature of this connection is by definition zero  and the torsion tensor is  given by
\begin{align}
 {T}^{\lambda}{}_{\mu \nu} = \Gamma^\lambda{ }_{\nu \mu} -\Gamma^\lambda{ }_{\mu \nu} = e_{A}{}^{\lambda}(\partial_{\mu} e^{A}{}_{\nu}-\partial_{\nu} e^{A}{}_{\mu}) ~.
\end{align}
We can relate the teleparallel linear connection to Christoffel symbols using the Ricci theorem
\begin{equation}\label{Ricci}
 \Gamma^\lambda_{\ \mu\nu}=\overset{\circ}{\Gamma}{}^\lambda_{\ \mu\nu} + \frac{1}{2} \left(
 T^{\ \lambda}_{\nu \ \mu} +T^{\ \lambda}_{\mu \ \nu} -T^{\lambda}_{\  \mu\nu} \right )~.
\end{equation}
Using \eqref{Ricci} it is straightforward to show that the Ricci scalar  can be 
written as
\begin{equation}
    \overset{\circ}{R}{}\equiv-T+B~,
\end{equation}
where we have defined the torsion scalar as
\begin{equation}
    \label{eq:Tscalardef}
    T = \frac{1}{4} T^{\rho}{}_{\mu 
\nu} T_{\rho}{}^{\mu \nu}+\frac{1}{2} T^{\rho}{}_{\mu \nu} T^{\nu 
\mu}{}_{\rho}-T^{\rho}{}_{\mu \rho} T^{\nu \mu}{}_{\nu} ~,
\end{equation}
and the boundary term given by $B=-2\overset{\circ}{\nabla}{}_\mu T^\mu$ and $T^\mu\equiv T^{\nu \mu  }{}_{\nu}$ is the vector  torsion .  

We can consider the Lagrangian given by the linear function of a torsion scalar 
\begin{equation}\label{actionTEGR}
    S_{\text{TEGR}}=-\frac{M_P^2}{2}\int \text{d}^4x \,e T~,
\end{equation}
where $e=\sqrt{-g}$. Due to the fact that the difference  between curvature and  
torsions scalars is just the total derivative term, the action 
\eqref{actionTEGR} is dynamically equivalent to the Einstein-Hilbert action and 
hence this theory is known as the teleparallel equivalent of general relativity 
(TEGR) \cite{Maluf:2013gaa,Aldrovandi:2013wha}.

Analogously to the $f(R)$ modifications of gravity,  we can then consider 
modifications of the teleparallel Lagrangian \eqref{actionTEGR} and consider 
theories with the action given by the  function of the torsion scalar 
\eqref{eq:Tscalardef}, i.e.
\begin{equation}\label{actionfT}
    S_{f(T)}=-\frac{M_P^2}{2}\int  \text{d}^4x \,e f(T)~,
\end{equation}
which is known as $f(T)$ gravity 
\cite{Ferraro:2006jd,Ferraro:2008ey,Bengochea:2008gz,Linder:2010py}.

A commonly supported viewpoint states that the theory generally propagates 5 DoFs \cite{Li:2011rn,Blagojevic:2020dyq}. 
However, a perturbation analysis shows the extra mode  appears only at fourth order or higher in  Minkowski and Friedmann-
Lema\^itre-Robertson-Walker (FLRW) backgrounds \cite{BeltranJimenez:2020fvy}, which is a sign of the strong coupling problem. For an analysis in the EFT framework, see \cite{Hu:2023juh,Hu:2023xcf}.

\subsection{A conformal equivalent of \texorpdfstring{$f(T)$}{f(T)} gravity and 
its EFT form} 
\label{Section:2.2}

The conformal equivalent of $f(T)$ gravity  can be derived along the lines of  the $f(R)$ gravity \cite{Li:2010cg,Yang:2010ji,Wright:2016ayu}. We introduce the auxiliary fields $\lambda$ and $\tilde{\Psi}$ and write the action \eqref{actionfT} as
\begin{equation}\label{actionfte}
    S_{f(T)}=-\frac{M_P^2}{2}\int  \text{d}^4x\, e  \Big[f(\tilde{\Psi})+\lambda(T-\tilde{\Psi})\Big]~.
\end{equation}
Variation with respect to $\lambda$  yields $\tilde{\Psi}=T$, and hence 
$\eqref{actionfte}$ is indeed dynamically equivalent to the original $f(T)$ 
action \eqref{actionfT}. Moreover, varying the action with respect to 
$\tilde{\Psi}$ yields $\lambda=f^{\prime}(\tilde{\Psi})$, where the prime 
represents the derivative with respect to $\tilde{\Psi}$. Then 
we insert the solution of $\lambda$ and rewrite the action in the  
Jordan frame
\begin{equation}\label{fT Jordan frame}
      S_{f(T)}=-\frac{M_P^2}{2}\int   \text{d}^4x \, e 
\Big[f^{\prime}(\tilde{\Psi})T+\Big(f(\tilde{\Psi})-\tilde{\Psi}f^{\prime}
(\tilde{\Psi})\Big)\Big]~.
\end{equation}
Imposing the conformal transformation  
$g_{\mu\nu}=\Omega^{2}(x)\hat{g}_{\mu\nu}$, which in tetrad formalism 
corresponds to transformations of cotetrads and the volume element  as $e^A_{\ 
\mu}=\Omega(x)\hat{e}^A_{\ \mu}$ and $e=\Omega^4(x)\hat{e}$ 
\cite{Maluf:2011kf,Yang:2010ji,Bamba:2013jqa}, and choosing the conformal factor 
to be $\Omega=(f^{\prime})^{-1/2}$, 
the action \eqref{fT Jordan frame} can be transformed into
\begin{equation}\label{fT conform1}
    S_{\text{conform}}=-\frac{M_P^2}{2}\int \text{d}^4x \,  \hat{e}  
\Big[\hat{T}+2(f^{\prime})^{-1}f^{\prime\prime}\nabla_\mu\tilde{\Psi}\hat{T}
^\mu-\frac{3}{2}(f^{\prime})^{-2}(f^{\prime\prime})^{2}\hat{g}^{\mu\nu}
\nabla_\mu\tilde{\Psi}\nabla_\nu\tilde{\Psi}+(f^{\prime})^{-2}(f-\tilde{\Psi}f^{
\prime})\Big]~.
\end{equation}
The hatted quantities  refer to conformally transformed ones. 
Then we  require $\Psi$ satisfying 
$\Psi^\prime(\tilde{\Psi})=\sqrt{3}f^{\prime\prime}/f^{\prime}$
in order to have the the kinetic term in \eqref{fT conform1} in the canonical  
form, which means $\Psi(\tilde{\Psi})=\sqrt{3}\ln f^{\prime}(\tilde{\Psi}),$. 
Finally, $f(T)$ gravity can be cast into a form
\begin{equation}\label{fT conform}
    S_{\text{conform}}=\frac{M_P^2}{2}\int  \hat{e} \text{d}^4x 
\Big[-\hat{T}-\frac{2}{\sqrt{3}}\hat{T}^\mu\nabla_\mu\Psi+\frac{1}{2}\hat{g}^{
\mu\nu}\nabla_\mu\Psi\nabla_\nu\Psi-V(\Psi)\Big]~,
\end{equation}
where
\begin{equation}
V(\Psi)=\Big(f^{\prime}(\Psi)\Big)^{-2}\Big(f(\Psi)-\tilde{\Psi}(\Psi)f^{\prime}
(\Psi)\Big)~.
\end{equation}

Unlike the case of $f(R)$ gravity, we are not able to recast the theory into a 
minimally coupled form due to the presence of the second term. This fact can be 
generalized to the case where disformal transformations are applied 
\cite{Hu:2023gui}. 
The first term is same as in  TEGR,  and hence would correspond to two 
propagating DoFs, but additional DoFs are introduced through the second and 
third terms. We can note also the wrong  kinetic term of the scalar field.
This conformal form of  the action allows us to understand the dynamics of the 
theory in a clearer way and is used often to analyse the dynamics of $f(T)$ 
gravity and other modified theories of gravity 
\cite{BeltranJimenez:2021auj,Hu:2023gui}.

The dynamics of  the auxiliary scalar field $\tilde{\Psi}$, introduced via 
conformal transformation, partially reflect the dynamics of $f(T)$ gravity. 
Therefore, both the tetrad fields and the conformal scalar field should be 
treated on equal footing, as they both contribute to the description of the 
gravity sector. This approach is straightforward within the torsional Effective 
Field Theory (EFT) framework, where these fields are incorporated into the 
torsional EFT action through fundamental geometric operators that respect the 
symmetries of the foliation structure, and the action can be written as 
\cite{Li:2018ixg,Hu:2023xcf}
\begin{align} \label{fTmodel}
 S =\int d^4x \sqrt{-g} \Big[  & \frac{M^2_P}{2} \Psi(t)R - \varLambda(t)  - 
b(t) g^{00} + \frac{M^2_P}{2} d(t) T^0 \Big] +S^{(2)}+\cdots~, 
\end{align}
where higher order sectors $S^{(2)}$ and beyond are used to denote all  the 
terms that explicitly start quadratic in perturbations. Note that the presence 
of the scalar field with
the coupling $T^{\mu}\nabla_{\mu}\Phi$ contributes to the operator $T^0$ in the 
EFT action.  We rewrite the Lagrangian density \eqref{fT conform} using the EFT 
approach and present its form in unitary gauge as follows
\begin{align}\label{conformalfTE}
   \mathcal{L}_{\text{conform}}& \overset{\text{u}}{=}\frac{M_{P}^{2}}{2}\hat{R}-\frac{M_{P}^{2}}{\sqrt{3}}\dot \Phi_{0}\hat{T}^{0}+\frac{M_{P}^{2}}{4}\dot \Phi_{0}^2g^{00}-\frac{M_{P}^{2}}{2}U(\Phi)~,
\end{align}
where $\Phi_0$ is the background value of the scalar field $\Phi$
and the dot symbol represents the  time
derivative.
It is easy to see that \eqref{conformalfTE}  is a special case of \eqref{fTmodel} with coefficients
\begin{align}\label{conformalfTE coefficients}
    \Psi(t)=1~,\quad b(t)=-\frac{M_{P}^{2}}{4}\dot  \Phi_{0}^2~,\quad 
d(t)=-\frac{2}{\sqrt{3}}\dot \Phi_{0}~,\quad\dot \Psi(t)=F(t)=0~,
\end{align}
and hence can be written as 
\begin{align} \label{model}
	S =\int d^4x \sqrt{-g} \Big[  & \frac{M^2_P}{2}  R - \varLambda(t) 
+\frac{3M^2_P}{16} d^2(t) g^{00} + \frac{M^2_P}{2} d(t) T^0 \Big] ~.
\end{align}

\section{The strong coupling of the \texorpdfstring{$f(T)$}{f(T)} conformal equivalent}\label{Section:3}

In this section we  focus on perturbation theory in the EFT form of the $f(T)$ 
conformal equivalent theory \eqref{fT conform}, in order to explore  the strong 
coupling problem. While the previous analysis shows this to second order scalar 
perturbations for $f(T)$ gravity \cite{Hu:2023juh} and the linear perturbations 
in its conformal equivalent \cite{Hu:2023xcf}, here we aim to show that the 
strong coupling is present in the same way between two conformal frames beyond 
the linear order.

\subsection{Perturbations of tetrads }

The tetrad fields are perturbed around the FLRW background
\begin{equation}
e_{\phantom{A}\mu}^{A}= 
\bar{e}^{A}_{\phantom{A}\mu}+e_{\phantom{(1)A}\mu}^{(1)A}+e_{\phantom{(2)A}\mu}^
{(2)A}+e_{\phantom{(3)A}\mu}^{(3)A}+\cdots,
\end{equation}
where the background tetrad is
\begin{equation}
        \bar{e}^{A}_{\phantom{A}\mu}=\left(\begin{array}{cc}
1 & 0\\
0 & a\delta_{i}^{a}
\end{array}\right)~,
\end{equation}
and the linear perturbations in the Newtonian gauge are
\begin{equation}       e_{\phantom{(1)A}\mu}^{(1)A}\equiv\left(\begin{array}{cc}
\phi & a\partial_{i}\chi\\
\partial^{a}\chi & -a\psi\delta_{i}^{a}
\end{array}\right)~, \label{basictetrad}
\end{equation}
where $\phi$, $\psi$, $\chi$ are the corresponding scalar perturbations, $a$ is the scale factor. At the linear level, we can define  
\begin{equation}
m_{\phantom{\nu}\mu}^{\nu}\coloneqq\bar{e}_{A}^{\phantom{A}\nu}e_{\phantom{(1)A}\mu}^{(1)A}-\delta_{\phantom{\nu}\mu}^{\nu}~.
\end{equation}
Beyond the linear level, we use the exponential ansatz \cite{Li:2018ixg,BeltranJimenez:2020fvy}\\
\begin{equation}
e_{\phantom{\nu}\mu}^{\nu}\coloneqq\left(e^{\bm{m}}\right)_{\phantom{\nu}\mu}^{\nu}\equiv\delta_{\phantom{\nu}\mu}^{\nu}+m_{\phantom{\nu}\mu}^{\nu}+\frac{1}{2}m_{\phantom{\nu}\rho}^{\nu}m_{\phantom{\nu}\mu}^{\rho}+\cdots~,
\label{Eexpand}
\end{equation}
and the  full tetrad is then obtained as $e^A{}_\mu=\bar{e}^A{}_\nu e^\nu_\mu$. The second order perturbations of the tetrad  are then expressed as
\begin{align}       e_{\phantom{(2)A}\mu}^{(2)A}&=\frac12\left(\begin{array}{cc}
\phi^2 
+\partial_i\chi \partial^i\chi
& a(\phi\partial_i\chi -\psi\partial_i\chi) \\ 
\phi\partial^a\chi - 
\psi\partial^a\chi 
& a(\psi^2\delta^a_i+ \partial_i \chi \partial^a \chi)
\end{array}\right)~,\label{tetradorder2}
\end{align}
and the cubic order  perturbations as
\begin{align}       
e_{\phantom{(3)A}\mu}^{(3)A}&=\frac16\left(\begin{array}{cc}
\phi^{3}+(2\phi-\psi)\,(\partial_i\chi)^{2}
& a\partial_{i}\chi(\phi^{2}
-\phi\psi+\psi^{2}+\partial_j\chi\partial^j\chi) \\ 
\partial^{a}\chi(\phi^{2}-\phi\psi+\psi^{2}
+\partial_j\chi\partial^j\chi) 
& -a\delta_{i}^{a}\psi^{3}+a(\phi-2\psi)\partial_{i}\chi\partial^{a}\chi
\end{array}\right)~.\label{tetradorder3}
\end{align}
We then find the metric perturbations up to the cubic order as follows
\begin{align}  
g_{00} &= -(1+2\phi+2\phi^2) -\frac{2}{3}\left[2\phi^{3}+(\phi+\psi)\,(\partial_i\chi)^{2}\right]+\cdots ~, \\ 
g_{0j} &=-a(\phi+\psi)\partial_j \chi -a (\phi^2-\psi^2)\partial_j\chi+\cdots~,\\
g_{ij} &= a^2\delta_{ij}(1-2\psi+2\psi^2) -\frac{2}{3}a^{2}\left[2\delta_{ij}\psi^{3}+(\phi+\psi)\,\partial_{i}\chi\partial_{j}\chi\right]+\cdots ~, \\  
g^{00} &= -(1-2\phi+2\phi^2) +\frac{2}{3}\left[2\phi^{3}+(\phi+\psi)\,(\partial_i\chi)^{2}\right]+\cdots ~, \\ 
g^{0j} &=-a^{-1}(\phi+\psi)\partial^{j} \chi +\frac{1}{a} (\phi^2-\psi^2)\partial^j\chi+\cdots~,\\
g^{ij} &= a^{-2}\delta_{ij}(1+2\psi+2\psi^2) +\frac{2}{3a^{2}}\left[2\delta^{ij}\psi^{3}+(\phi+\psi)\,\partial^{i}\chi\partial^{j}\chi\right] +\cdots~.
\end{align}
It should be noted that before actually applying perturbation methods in the Newtonian gauge, we have to recover the full spacetime diffeomorphism invariance of EFT formalism by St\"{u}ckelberg trick in order to avoid excessive gauge fixing. After we perform a time coordinate transformation $t\to t+\pi$, the scalar mode $\pi$ as the Goldstone mode will become explicit. 
We provide a simple demonstration of this procedure through several examples. All the background time-dependent scalar operators, such as $\varLambda(t)$, should be treated in the new time coordinate and expanded with orders of $\pi$ as
\begin{align}
\varLambda(t)\rightarrow\varLambda(t+\pi)=\varLambda(t)+\dot\varLambda(t)\pi+\frac{1}{2}\ddot\varLambda(t)\pi^2+\cdots~,
\end{align}
in which $\pi$ appears not derivated. And operators with temporal components transform under time coordinate transformation and generate terms with derivatives acting on $\pi$, for example
\begin{align}
T^{0}\rightarrow T^{0}+T^{\mu}\partial_{\mu}\pi~.
\end{align}

\subsection{Background level}

At the background level, we obtain two independent background evolution equations
\begin{align}
 \dot{H} &= \frac{3}{16}d^2(t)  -\frac{1}{4}\dot{d}(t) +\frac{3}{4}Hd(t)    ~, \label{H} \\
 \varLambda &= M_{P}^2 \Big( 3H^2 + \dot{H} + \frac{1}{4}\dot{d}(t) +\frac{3}{4}Hd(t) \Big) ~. 
\end{align}
We simplify the expression of $\varLambda$ through the first equation \eqref{H} as 
\begin{equation}\label{Lambda}
\varLambda = \frac{3}{16}M_{P}^2 \Big( d^2(t) +8Hd(t) + 16H^2 \Big) ~.
\end{equation}
In the following, we replace all the $\dot{H}$ and $\varLambda$ and their time evolution by equations \eqref{H} and \eqref{Lambda}.

\subsection{Linear order scalar perturbations}\label{Section:3.3}

In this subsection, we will review the results of the linear scalar perturbations of $f(T)$ conformal equivalent theory in EFT approach \cite{Hu:2023xcf}.
Varying the action with respect to perturbation variables we obtain the corresponding equations of motion, i.e., $\mathcal{E}^{(1)}_{I}=0,\ I=1,...,4$, where we denote the field indices $1,2,3,4$ as $\pi,\psi,\phi$ and $\chi$ respectively, for simplicity.
The equations of motion are given by
\begin{align}
\mathcal{E}^{(1)}_{A}=W^{(1)}_{AB}\ddot q_{B}+K^{(1)}_{A}~,
\end{align}
where
\begin{equation}
    W^{(1)}_{AB}=
    \begin{pmatrix}
    \mathcal{J}_{\{\ddot \pi,\ddot \psi\}} & \mathcal{O}\\
    \mathcal{O} & \mathcal{O}
\end{pmatrix}~
\end{equation}
with the $2\times2$ matrix $\mathcal{J}_{\{\ddot \pi,\ddot \psi\}}$ and the zero matrix $\mathcal{O}$, and $K^{(1)}_{A}$ are the rest part of the equations linearly depending on $\dot q_A$ and $q_A$. We list their specific expression in Appendix \ref{appendixA}.
The Jacobian of $\{\ddot \pi,\ddot \psi\}$
\begin{equation}
    \mathcal{J}_{\{\ddot \pi,\ddot \psi\}}=
    \begin{pmatrix}
       \mathcal{W}_{\ddot{\pi}}^{1} & \mathcal{W}_{\ddot{\psi}}^{1}\\
       \mathcal{W}_{\ddot{\pi}}^{2} & \mathcal{W}_{\ddot{\psi}}^{2}
     \end{pmatrix}=
     \begin{pmatrix}
         \frac{3}{8}M_{P}^{2}a^{3}d^{2}(t) & -\frac{3}{2}M_{P}^{2}a^{3}d(t)\\
        -\frac{3}{2}M_{p}^{2}a^{3}d(t) & 6M_{p}^{2}a^{3}
      \end{pmatrix}~,
\end{equation}
generated from the  equations of motion, its structure is naturally degenerated 
due to special coefficient combinations in the action \eqref{model}, as 
mentioned in our previous work \cite{Hu:2023xcf}. The degeneracy of the Jacobian 
implies that there is at least one further constraint 
$C_1(\pi,\psi,\chi,\phi;\dot{\pi},\dot{\psi};\partial^2) = 0$, where 
$\partial^2=\partial_i\partial^i$. 
The explicit expression for this constraint can be found in Appendix \ref{appendixA}, where it is derived as a linear combination of the first two equations of motion.

Meanwhile, there are two primary  constraints generated by variations. One of 
the primary constraint is obtained through the equation of motion with respect 
to $\phi$, i.e., 
$\mathcal{E}^{(1)}_3(\pi,\psi,\phi;\dot\pi,\dot\psi;\partial^2)=0$. 
We can also easily obtain another constraint from the variation with respect to $\chi$ as
\begin{equation} 
    \mathcal{E}^{(1)}_{4}=d(t)(H\pi+\psi)=0~.
\end{equation}
In general, $d(t)$ is not equal to zero,  since $d(t)=0$ would be a trivial 
solution, leading to an action equivalent to the Einstein-Hilbert case. Thus, we 
focus on the solution 
\begin{equation}
    \psi=-H\pi~.\label{psisolution}
\end{equation}
Both constraints $C_1=0$ and $\mathcal{E}^{(1)}_3=0$  would be further 
simplified after taking Eq.\eqref{psisolution} into consideration.
Then $\chi$ and $\phi$ can be solved and expressed by $\pi$ as
\begin{align}
 \chi&= \frac{\pi}{a}~,\label{chisolution} \\ 
 \phi&= \dot \pi-\frac{4}{3a^2\left(d(t)+4 H\right)}\partial^2\pi~,\label{phisolution}
\end{align}
with the assumption that both $d(t)$ and  the denominator $d(t)+4 H$ do not 
vanish. Note that an additional constraint, namely
$C_2(\pi;\dot\pi;\partial^2)=\partial^4 \pi=0$, is obtained through  $\pi$ 
variation  after substituting all the solutions above into the action. 
 Then the results imply that no propagating scalar 
mode exists in linear perturbation.

\subsection{Second-order scalar perturbations}

In this subsection  we focus on the action at the cubic order in  perturbations. 
Since the number of terms drastically increases in the cubic action, it is not 
necessary nor useful to show all of them, and we show only those that are of 
interest for the purpose of this work. After substituting all   solutions 
obtained at the linear order, we can write the cubic action schematically as
\begin{equation}       
\mathcal{L}^{\text{kin}}_{3}=\mathcal{L}^{\text{kin}}_{3,t3}
+\mathcal{L}^{\text{kin}}_{3,t2}
+\mathcal{L}^{\text{kin}}_{3,t1}+\mathcal{L}^{\text{kin}}_{3,t0}~,
\end{equation}
where the notation $tn, n=3,...,0$ stands for the total number of time derivatives contained in the given term.
The term containing three time derivatives  in the Lagrangian  is given by
\begin{equation}
    \mathcal{L}^{\text{kin}}_{3,t3}\simeq
    -\frac{1}{12}M_{p}^{2}a\Big(54a^{2}H^{2}\dot{\pi}^{3}+6\dot{\pi}^{2}\partial^{2}\dot{\pi}+13\partial^{i}\pi\partial_{i}\dot{\pi}\ddot{\pi}+25\dot{\pi}\partial_{i}\dot{\pi}\partial^{i}\dot{\pi}+108a^{2}H^{2}\pi\dot{\pi}\ddot{\pi}+13\dot{\pi}\partial^{i}\pi\partial_{i}\ddot{\pi}\Big)~.
\end{equation}
Performing integration by parts and using the background evolution equations, 
the order of the time derivatives can be reduced to the second order in time 
derivatives 
\begin{equation}\label{third order time derivatives in cubic}
    \mathcal{L}^{\text{kin}}_{3,t3}\simeq\frac{9}{16}M_{p}^{2}a^{3}H\Big(-4\dot{d}(t)+3d^{2}+12Hd(t)+24H^{2}\Big)\dot{\pi}^{2}\pi-\frac{13}{24}M_{p}^{2}aH\dot{\pi}^{2}\partial^{2}\pi~.
\end{equation}
Similarly, the term $\mathcal{L}^{\text{kin}}_{3,t2}$ can be reduced to 
\begin{align}\label{second order time derivatives in cubic}
    \mathcal{L}^{\text{kin}}_{3,t2}\simeq
    &\frac{9}{16}M_{p}^{2}a^{3}H\Big(4\dot{d}(t)-3a^{2}d^{2}(t)-12Hd(t)-24H^{2}\Big)\dot{\pi}^{2}\pi+\frac{13}{24}M_{p}^{2}aH\dot{\pi}^{2}\partial^{2}\pi\\ \nonumber
    &\ \ \ \ 
+\frac{1}{384}M_{p}^{2}a\Bigl(-4\dot{d}(t)+3d^{2}(t)+12Hd(t)+16H^{2}\Bigr)\dot{
\pi}\partial_{i}\pi\partial^{i}\pi~.
\end{align}
  We can see that Eq.\eqref{third order time derivatives in cubic} and the first 
term of Eq.\eqref{second order time derivatives in cubic} cancel each other out, 
resulting in the elimination of all higher-order time derivatives, leaving only 
some linear first-order time derivative terms
\begin{equation}\label{third and second order time derivatives in cubic}
\mathcal{L}^{\text{kin}}_{3}\simeq 
\frac{1}{384} M_P^2 a \left(-4 \dot d(t)+3 d^2(t)+12Hd(t)+16 H^2\right)\dot\pi\partial_i \pi \partial^i \pi+\mathcal{L}^{\text{kin}}_{3,t1}+\mathcal{L}^{\text{kin}}_{3,t0} ~.
\end{equation}

We conclude that no propagating  scalar mode appears in the  second-order 
perturbative analysis of the $f(T)$ conformal theory, and hence we demonstrate 
that the strong coupling issue remains in the conformal equivalent case.

\section{The strong coupling of a modified \texorpdfstring{$f(T)$}{f(T)} 
conformal theory} \label{Section:4}

In this section, we introduce a method to find new modified teleparallel 
theories  of gravity within the EFT approach, where the order at which strong 
coupling appears is lowered. A simple example of such a theory is one where we 
modify the coupling of the operator $T^0$ by changing the relation between 
coefficients $b$ and $d$ described by equations \eqref{conformalfTE 
coefficients} in the EFT action \eqref{conformalfTE} by introducing a parameter 
$\epsilon$, which quantifies the deviation from the $f(T)$ equivalent case, i.e.
\begin{align}\label{dtprime}
      d\to d^\prime(t)=d(t)+\epsilon\, c(t)~.
\end{align}
The correspondingly modified action takes the form
\begin{align} \label{model1}
 S =\int d^4x\, e \Big[  & \frac{M^2_P}{2} R - \varLambda(t) +\frac{3M^2_P}{16} d^2(t) g^{00} + \frac{M^2_P}{2} \Big(d(t)+\epsilon\, c(t)\Big) T^0 \Big] ~,
\end{align}
which can include the covariant form of the action
\begin{equation}
    S=\frac{M_{P}^{2}}{2}\int\text{d}^{4}x\, e \Big[-T-V(\Psi)+\frac{1}{2}g^{\mu\nu}\nabla_{\mu}\Psi\nabla_{\nu}\Psi+(-\frac{2}{\sqrt{3}}+\epsilon C(\Psi))T^{\mu}\nabla_{\mu}\Psi\Big]~
\end{equation}
as a specific example.
This specific modified action  can be recast into the following form through a 
inversely conformal transformation and some manipulations 
\begin{equation}
    S=\frac{M_{P}^{2}}{2}\int \text{d}^{4}x\, e\Big[-F(\tilde\Phi)T-\frac{1}{2}\nabla^{\mu}\tilde\Phi\nabla_{\mu}\tilde\Phi-\tilde V(\tilde\Phi)\Big]~,
\end{equation}
where the details are presented in Appendix \ref{appendixB}. The above action with a non-minimal coupling between a scalar field and the torsion scalar which is referred to as teleparallel dark energy. Especially, the model with a specific choice as $F(\tilde\phi)=1+\xi\tilde\phi^2$ were originally discussed in Ref. \cite{Geng:2011aj,Wei:2011yr,Geng:2011ka} and more generally extended couplings are considered in Ref. \cite{Otalora:2013tba,Dil:2015eum,Bahamonde:2018miw}.

Following the same approach as   in Section~\ref{Section:3}, we need to utilize 
the gauge recovering procedure to ensure that we can correctly fixing scalar 
gauge in the following perturbation procedure. Then we insert the perturbed 
basic variables in Newtonian gauge into the action and expand the action order 
by order around the FLRW background.

\subsection{Background level}

At the background level   we obtain the equations of motion of $\dot H$ and 
$\varLambda$ as
\begin{align}
\dot H &= \frac{1}{16}\left(3 d^2(t)+12Hd(t)-4 \dot d(t)+12 \epsilon\, c(t) H-4 \epsilon\, \dot c(t)\right)~,\\
\varLambda &= \frac{3}{16}M_P^2\Big( d^2(t)+8 Hd(t)+16 H^2+8 \epsilon\, c(t) H \Big)~.
\end{align}
 In the following we  use the above equations as background solutions to 
simplify terms including $\dot H$ and $\varLambda$ and their time evolution.

\subsection{Linear order scalar perturbations}
We derive the equations of motion from the modified Lagrangian
\begin{align}
\mathcal{\tilde E}^{(1)}_{A}=\tilde W^{(1)}_{AB}\ddot q_{B}+\tilde K^{(1)}_{A}~,
\end{align}
where
\begin{equation}
    \tilde W^{(1)}_{AB}=
    \begin{pmatrix}
    \mathcal{\tilde J}_{\{\ddot \pi,\ddot \psi\}} & \mathcal{O}\\
    \mathcal{O} & \mathcal{O}
\end{pmatrix}~,
\end{equation}
with the $2\times2$ matrix  $\mathcal{\tilde J}_{\{\ddot \pi,\ddot \psi\}}$ and 
$\tilde 
K^{(1)}_{A}$ are the remaining parts of the equations linearly depending on $\dot q_A$ and $q_A$. Their specific expressions are written in Appendix \ref{appendixA}. We also use the tilde symbol ``$\tilde{\phantom{A}}$" to represent the modified action \eqref{model1}.

Since the  modification of coefficient $d$ we introduced breaks the initial 
degeneracy of the Jacobian $\mathcal{\tilde J}_{\{\ddot \pi,\ddot \psi\}}$, 
there is no analogue to the constraint  previously denoted as $C_1$, which could 
 be  deduced from the equations of motion with respect to $\pi$ and $\psi$. That 
means we only have two primary constraints, which correspond respectively to 
variations with respect to $\chi$ and $\phi$. 

The first constraint is
\begin{equation}\label{constraint3}
    \mathcal{\tilde E}^{(1)}_{4}=-M_P^2 a^2\Big(\epsilon\, c(t)+d(t)\Big)\Big(H \pi+\psi\Big)=0~,
\end{equation}
which can be solved either as $\epsilon\, c(t)+d(t)=0$, corresponding to $f(R)$ gravity in the Einstein frame,  or by $\psi=-H\pi$. The second constraint is $\mathcal{\tilde E}^{(1)}_{3}=0$, which after using $\psi=-H\pi$ and its time evolution, leads to a  solution for $\phi$ as

\begin{equation}
\phi=\dot \pi -\frac{4}{3 a^2}\frac{ d(t)+4 H+ \epsilon\, c(t)}{\Big(d(t)+4 H\Big)^2+8 \epsilon Hc(t)}\partial^2 \pi~.
\end{equation}
Since $\psi$ and $\pi$ are related by the constraint \eqref{constraint3}, for example when we replace $\psi$ in terms of $\pi$ in the action, the Jacobian $\mathcal{\tilde J}_{\{\ddot \pi,\ddot \psi\}}$ is again degenerated and their equations of motion $\mathcal{\tilde E}^{(1)}_{1}=0,\ A=1,2$ become relevant. Further constraint should be derived by linear combinations of their equations of motion as $\tilde C_{}(\pi,\chi; \dot \pi;\partial^2)= 0$.

After some  algebra, we can express $\chi$ in terms of $\pi$ from the 
constraint $\tilde C_{}$ as
\begin{align}
\chi=\mathcal{\tilde C}_{\dot{\pi}}\dot{\pi}+\mathcal{\tilde C}_{\pi}\pi~.
\end{align}
Hence, replacing these scalar perturbation with their solutions  in the action, 
the kinetic term of $\pi$ gets cancelled and hence we do not find any 
propagating scalar modes at the linear order. 

We provide a complete  form of solutions for scalar perturbations in Appendix 
\ref{appendixA} since they are needed in a subsequent analysis of second-order 
scalar perturbations. 
It can be clearly seen that when $\epsilon$ approaches  zero the solutions of 
$\psi$, $\phi$ and $\chi$ will revert to solutions found in the conformal 
equivalent form of $f(T)$ gravity in Sec. \ref{Section:3.3}.

\subsection{Second-order scalar perturbations}

We can now proceed to the second-order perturbations,  where we follow the 
steps performed in Sec.~\ref{Section:3.3}: we  insert the solutions 
from the linear order into the cubic action, and we perform integration by 
parts to lower the order of time derivatives whenever possible. While the full 
cubic  action is     lengthy, here we show only the relevant third-order part 
in time derivatives, namely
\begin{align} \label{L3 kin}
\mathcal{\tilde L}_{3,t3}^{\text{kin}}\simeq\mathcal{D} 
\Big\{\hat{\mathcal{C}}_{\partial^{i}\dot{\pi}\partial^{j}\dot{\pi}\partial_{j}
\partial_{i}\dot{\pi}}^{\text{M}}\partial^{i}\dot{\pi}\partial^{j}\dot{\pi}
\partial_{j}\partial_{i}\dot{\pi}-\hat{\mathcal{C}}_{\dot{\pi}^{2}\dot{\partial}
^{4}\pi}^{\text{M}}\left[-\dot{\pi}^{2}\dot{\partial}^{4}\pi+2\partial_{j}
\partial_{i}\dot{\pi}(\partial^{i}\dot{\pi}\partial^{j}\dot{\pi}+\dot{\pi}
\partial^{j}\partial^{i}\dot{\pi})\right] \Big\} ~,
\end{align}
where the coefficient functions are
\begin{align}
&\hat{\mathcal{C}}_{\partial^{i}\dot{\pi}\partial^{j}\dot{\pi}\partial_{j}
\partial_{i}\dot{\pi}}^{\text{M}}=2\epsilon d(t)\Big(2d(t)+\epsilon 
c(t)\Big)c(t)~,\\
&\hat{\mathcal{C}}_{\dot{\pi}^{2}\dot{\partial}^{4}\pi}^{\text{M}}=-\frac{1}{3}\Big(d^{2}(t)+8Hd(t)+16H^{2}+8\epsilon Hc(t)\Big)~,
\end{align}
with
\begin{align}
    \mathcal{D}&=\frac{2M_{P}^{2}\epsilon^{2}\Big(2d(t)+\epsilon c(t)\Big)^{2}c^{2}(t)}{a\Big(d^{2}(t)+8Hd(t)+16H^{2}+8\epsilon Hc(t)+2H\Big)^{3}\Big(d(t)+\epsilon c(t)\Big)^{2}}~.
\end{align}
We mention that the terms in the bracket of the second term in  Eq.\eqref{L3 kin} 
is a combination of the following expression
\begin{align}
    &-\dot \pi^2  \partial^4 \dot\pi+2  \partial_j  \partial_i \dot\pi( 
\partial^i \dot\pi  \partial^j \dot\pi+\dot \pi  \partial^j \partial^i \dot\pi)
   =\partial_i\Big[\partial_j(\dot \pi^2  \partial^i\partial^j \dot\pi)-2\dot  
\pi^2  \partial^i\partial^2 \dot\pi\big]~.
\end{align}
This is  a total spatial derivative, which can be handled as a boundary term in 
a homogeneous and isotropic background. Thus, all terms proportional to it in 
the cubic action can be neglected. After simplification, some terms with 
total third-order time derivatives remain, expressed by the total third-order 
time derivatives left, which are given by
\begin{equation} \label{cubicLagrangian}
\mathcal{\tilde L}^{\text{kin}}_{3,t3}  \simeq 
\hat{\mathcal{C}}_{\partial^{i}\dot{\pi}\partial^{j}\dot{\pi}\partial_{j}
\partial_{i}\dot{\pi}}^{\text{M}}\partial^{i}\dot{\pi}\partial^{j}\dot{\pi}
\partial_{j}\partial_{i}\dot{\pi}~,
\end{equation}
whose variation with respect to $\pi$  reveals equations of motion with 
second-order time derivatives. The entire kinetic part in the effective cubic 
action is generated from the modification. This implies that by introducing 
additional coupling terms, we lower the order of the kinetic terms and find the 
propagating scalar mode in the second-order perturbation. Note that this 
feature is consistent with the results of Section~\ref{Section:3}, since $\mathcal{\tilde L}^{\text{kin}}_{3,t3}$ vanishes in the limit $\epsilon\rightarrow 0$, and is 
one of the main results of the present work.

\section{Conclusions}\label{Section:5}

The strong coupling problem refers to the case where a theory has a 
different number of DoFs perturbatively and 
non-perturbatively. In the case of $f(T)$ gravity, it is known that the theory 
has only two propagating DoFs at the  level of linear perturbations around 
Minkowski and FLRW backgrounds, however at the non-perturbative level it is 
expected to have 5 DoFs. Hence a  question arises naturally, namely at which 
level of perturbation theory these new DoFs appear. 

The previous results  around Minkowski background suggested that the new DoFs of $f(T)$ gravity become dynamical only up to the level of the quartic action \cite{BeltranJimenez:2020fvy} (considering only Lorentz perturbations). 
Additionally, the strong coupling mode beyond cubic order was suggested to appear in the FLRW case too, according to the EFT approach of $f(T)$ gravity \cite{Hu:2023juh}, which we have further explored in this paper. 
However, a complete understanding of the strong coupling problem likely requires a full perturbative analysis of both Lorentz and metric perturbations at the quartic level. 
Unfortunately, such an analysis is highly complex and currently beyond our capabilities.

We started by  considering the conformal equivalent form of $f(T)$ gravity,  in which case the corresponding EFT form is constructed non-perturbatively by linear operators. As we showed,   the scalar 
perturbations are not dynamical up to   cubic action, and we expect them to 
appear at the quartic action, similar to the Minkowski case.  This result   
agrees with the previous analysis using the EFT approach in $f(T)$ gravity 
\cite{Hu:2023juh,Hu:2023xcf}, however here it is demonstrated explicitly
through the use of the conformal equivalent form of   $f(T)$ gravity. This 
approach demonstrates manifestly the dynamical equivalence of the conformal 
equivalent form of $f(T)$ gravity, as expected.

We proceeded by  exploring whether it is possible to lower the order of 
perturbation theory at which the new DoFs appear, and we constructed a 
modification of the EFT form of the conformal equivalent of $f(T)$ gravity. We 
  considered a rather simple and natural modification where the coefficient 
of the coupling of $T^0$ operator is changed from $d$ to 
$d^\prime(t)=d(t)+\epsilon\, c(t)$, which corresponds to a new theory of gravity 
distinct from the $f(T)$ class, nevertheless still constructed by linear 
operators. Finally, we have  confirmed that in this model the new 
propagating DoF appears at the level of the cubic action, and hence the order 
at which this new mode appears in the modified case  is effectively lowered 
compared to the original $f(T)$ case.

Our work  opens up a new avenue to address the issue of strong coupling in 
modified teleparallel gravity theories, applying the EFT framework  which 
offers a better control at which order of perturbations the new DoFs appear. 
This approach can help us  to understand more clearly the strong coupling issue 
in modified teleparallel gravity, as it allows for its investigation at lower 
orders of perturbation theory, where the problem becomes more manageable. 
In order to proceed to  simpler cases, where the problem becomes more 
tractable, one option would be to consider the New General Relativity theory, 
where the coefficients in the torsion scalar \eqref{eq:Tscalardef} differ from 
the TEGR case. Thus, the strong coupling problem is easier to be analyzed 
comparing to $f(T)$ gravity \cite{BeltranJimenez:2019tme}. Another option would 
be to explore theories more closely related to $f(T)$ gravity, such as the one 
presented in Section~\ref{Section:4}. Interestingly, the strongly coupled modes 
appear at the cubic action level in both cases, which naturally motivates 
further investigation of the relationship between them, a direction we plan to 
pursue in future work.

Lastly, we believe that our results may contribute to a new classification 
scheme for modified teleparallel gravity theories, based on the order at which 
the strong coupling problem arises within the EFT framework. This 
classification could help to identify all modified teleparallel theories free 
from the strong coupling issue, as well as theories where the problem appears 
at lower orders in perturbation theory, offering valuable insights into the 
nature of the strong coupling issue.

\acknowledgments

We thank Masahide Yamaguchi, Erik Jensko, Xinchen He, Xin Ren and Ya-Qi Zhao for valuable 
discussions. 
This work is supported  in part by National Key R\&D Program of China 
(2021YFC2203100), by National Natural Science Foundation of China (NSFC) 
(12261131497,
12347137,12433002), by CAS young  interdisciplinary innovation team 
(JCTD-2022-20), by China Postdoctoral Science Foundation (2024M753076), by 111 
Project (B23042), by Fundamental Research Funds for Central Universities, by CSC 
Innovation Talent Funds, by USTC Fellowship for International Cooperation, by 
USTC Research Funds of the Double First-Class Initiative.  MK is supported 
through  SASPRO2 project \textit{AGE of Gravity: Alternative Geometries of 
Gravity}, which has received funding from the European Union's Horizon 2020 
research and innovation programme under the Marie Skłodowska-Curie grant 
agreement No. 945478.
MK and ENS acknowledge the contribution of the     COST Action CA23130 
“Bridging high and low energies in search of quantum gravity (BridgeQG)”.
We acknowledge the use of the clusters LINDA \& JUDY of 
the particle cosmology group at USTC.

\appendix
\section{Explicit expressions of the linear scalar perturbations}\label{appendixA}
\subsection{\texorpdfstring{$f(T)$}{f(T)} conformal equivalent formalism}

The equations of motion of the $f(T)$ conformal equivalent formalism are given by the following form
\begin{align}
\mathcal{E}^{(1)}_{A}=W^{(1)}_{AB}\ddot q_{B}+K^{(1)}_{A}~,
\end{align}
where
\begin{equation}
    W^{(1)}_{AB}=
    \begin{pmatrix}
    \mathcal{W}_{\ddot{\pi}}^{1} & \mathcal{W}_{\ddot{\psi}}^{1} & 
 0 & 0\\
     \mathcal{W}_{\ddot{\pi}}^{2} & \mathcal{W}_{\ddot{\psi}}^{2} & 0 & 0\\
     0 & 0 & 0 & 0\\
     0 & 0 & 0 & 0
    \end{pmatrix}~,    
\end{equation}
and
\begin{equation}
    K^{(1)}_{A}=
    \begin{pmatrix}
    \mathcal{K}_{\pi}^{1} & \mathcal{K}_{\psi}^{1} & \mathcal{K}_{\phi}^{1} & \mathcal{K}_{\chi}^{1}\\
\mathcal{K}_{\pi}^{2} & \mathcal{K}_{\psi}^{2} & \mathcal{K}_{\phi}^{2} & \mathcal{K}_{\chi}^{2}\\
\mathcal{K}_{\pi}^{3} & \mathcal{K}_{\psi}^{3} & \mathcal{K}_{\phi}^{3} & 0\\
\mathcal{K}_{\pi}^{4} & \mathcal{K}_{\psi}^{4} & 0 & 0
\end{pmatrix}
\begin{pmatrix}\pi\\
\psi\\
\phi\\
\chi
\end{pmatrix}~,
\end{equation}
with the following coefficients
\begin{align}
    \mathcal{W}_{\ddot{\pi}}^{1}&=\frac{3}{8}M_{P}^{2}a^{3}d^{2}(t)~,
    &\mathcal{W}_{\ddot{\psi}}^{1}&=-\frac{3}{2}M_{P}^{2}a^{3}d(t)~,\\
    \mathcal{W}_{\ddot{\pi}}^{2}&=-\frac{3}{2}M_{p}^{2}a^{3}d(t)~,
    &\mathcal{W}_{\ddot{\psi}}^{2}&=6M_{p}^{2}a^{3}~,
\end{align}
\begin{align}
   \mathcal{K}^1_{\pi}&= 
\frac{3}{8}M_{P}^{2}a^{3}\Big(3Hd(t)+2\dot{d}(t)\Big)d(t)\,\partial_{t}-\frac{3}
{8}M_{P}^{2}ad^{2}(t)\partial^{2} \nonumber \\
   &\ \ \ \ -\frac{3}{64}M_{P}^{2}a^{3}\Big(-24H\dot{d}(t)-8\ddot{d}(t)+9d^{3}(t)+72Hd^{2}(t)+144H^{2}d(t)\Big)d(t)~,\\
   \mathcal{K}^1_{\psi}&=-\frac{9}{8}M_{P}^{2}a^{3}\Big(d(t)+8H\Big)d(t)\,\partial_{t}+M_{P}^{2}ad(t)\partial^{2}~,\\
   \mathcal{K}^1_{\phi}&=-\frac{3}{8}M_{P}^{2}a^{3}\Big(d(t)+ 
4H\Big)d(t)\,\partial_{t}-\frac{1}{2}M_{P}^{2}ad(t)\partial^{2}-\frac{9}{16}M_{P
}^{2}a^{3}\Big(d^{2}(t)+8Hd(t)+16H^{2}\Big)d(t)~,\\
   \mathcal{K}^1_{\chi}&=-M_{P}^{2}a^{2}Hd(t)\partial^{2}~,
\end{align}
\begin{align}
   \mathcal{K}^2_{\pi}&= 
\frac{3}{8}M_{p}^{2}a^{3} 
\Bigl(-8\dot{d}(t)+3d^{2}(t)\Big)\,\partial_{t}+M_{p}^{ 2}ad(t)\partial^{2} 
\nonumber  \\    &\ \ \ \ 
+\frac{3}{32}M_{p}^{2}a^{3}\Big(-16\ddot{d}(t)+12d(t)\dot{d}(t)+9d^{3}(t)+72Hd^{
2}(t)+144H^{2}d(t)\Big)~,\\
   \mathcal{K}^2_{\psi}&=18M_{p}^{2}a^{3}H\,\partial_{t}-2M_{p}^{2}a\partial^{2}~,\\
   \mathcal{K}^2_{\phi}&=\frac{3}{2}M_{p}^{2}a^{3}\Bigl(d(t)+4H\Big)\,\partial_{t}+2M_{p}^{2}a\partial^{2}+\frac{9}{8}M_{p}^{2}a^{3}\Big(d^{2}(t)+8Hd(t)+16H^{2}\Big)~,\\
   \mathcal{K}^2_{\chi}&=-M_{p}^{2}a^{2}d(t)\partial^{2}~,
\end{align}
\begin{align}
\mathcal{K}^3_{\pi}&=\frac{3}{8}M_{P}^{2}a^{3}\Big(d(t)+4H\Big)d(t)\,\partial_{
t }-\frac{1}{2}M_{P}^{2}ad(t)\partial^{2}\nonumber  \\  
   &\ \ \ \ +\frac{3}{32}M_{P}^{2}a^{3}\Big(4d(t)\dot{d}(t)+16H\dot{d}(t)-3d^{3}(t)-24Hd^{2}(t)-48H^{2}d(t)\Big)~,\\
   \mathcal{K}^3_{\psi}&=-\frac{3}{2}M_{P}^{2}a^{3}\Big(d(t)+4H\Big)\,\partial_{t}+2M_{P}^{2}a\partial^{2}~,\\
   \mathcal{K}^3_{\phi}&=-\frac{3}{8}M_{p}^{2}a^{3}\Big(d^{2}(t)+8Hd(t)+16H^{2}\Big)~,
\end{align}
and
\begin{align}
   \mathcal{K}^4_{\pi}&=-M_{p}^{2}a^{2}d(t)H\partial^{2}~,&\mathcal{K}^4_{\psi}&=-M_{p}^{2}a^{2}d(t)\partial^{2}~.
\end{align}
The degeneracy of the non-zero submatrix of $W^{(1)}_{AB}$ implies a further constraint $C_1\coloneqq C_{1A}q_A= 0$ where
\begin{align}  \label{constraintconformal} 
     C_{1A}=\begin{pmatrix}\mathcal{C}_{\pi}^{1} & \mathcal{C}_{\phi}^{1} & 
\mathcal{C}_{\psi}^{1} & \mathcal{C}_{\chi}^{1}\end{pmatrix} ,
\end{align}
with coefficients
\begin{align}     
\mathcal{C}_{\pi}^{1}&=\frac{9}{8}M_{P}^{2}a^{3}\Big(d(t)+4H\Big)d^{2}(t)\,
\partial_{t}-\frac{1}{2}M_{P}^{2}ad^{2}(t)\partial^{2}\nonumber \\ 
    &\ \ \ \ -\frac{9}{32}M_{P}^{2}a^{3}\Big(-4d(t)\dot{d}(t)-16H\dot{d}(t)+3d^{3}(t)+24Hd^{2}(t)+48H^{2}d(t)\Big)d(t)~,
    \\
    \mathcal{C}_{\phi}^{1}&=-\frac{9}{8}M_{P}^{2}a\Big(a^{2}d^{2}(t)\phi+8a^{2}Hd(t)\phi+16a^{2}H^{2}\Big)d(t)~,\\
    \mathcal{C}_{\psi}^{1}&=-\frac{9}{2}M_{P}^{2}a^{3}\Big(d(t)+4H\Big)d(t)\,\partial_{t}+2M_{P}^{2}ad(t)\partial^{2}~,\\
    \mathcal{C}_{\chi}^{1}&=-M_{P}^{2}a^{2}\Big(d(t)+4H\Big)d(t)\partial^{2}~.
\end{align}

\subsection{The modified \texorpdfstring{$f(T)$}{f(T)} conformal theory}

The equations of motion in the case of modified \texorpdfstring{$f(T)$}{f(T)} 
conformal theory formalism are given by  
\begin{align}
\mathcal{\tilde E}^{(1)}_{A}=\tilde W^{(1)}_{AB}\ddot q_{B}+\tilde K^{(1)}_{A}~,
\end{align}
where
\begin{equation}
    \tilde W^{(1)}_{AB}=
    \begin{pmatrix}
    \mathcal{\tilde{W}}_{\ddot{\pi}}^{1} & \mathcal{\tilde{W}}_{\ddot{\psi}}^{1} & 0 & 0\\
    \mathcal{\tilde{W}}_{\ddot{\pi}}^{2} & \mathcal{\tilde{W}}_{\ddot{\psi}}^{2} & 0 & 0\\
    0 & 0 & 0 & 0\\
    0 & 0 & 0 & 0
\end{pmatrix}~,    
\end{equation}
and
\begin{equation}
    \tilde K^{(1)}_{A}=
    \begin{pmatrix}
    \mathcal{\tilde{K}}_{\pi}^{1}  & \mathcal{\tilde{K}}_{\psi}^{1} & 
\mathcal{\tilde{K}}_{\phi}^{1} & \mathcal{\tilde{K}}_{\chi}^{1}\\
\mathcal{\tilde{K}}_{\pi}^{2} & \mathcal{\tilde{K}}_{\psi}^{2} & \mathcal{\tilde{K}}_{\phi}^{2} & \mathcal{\tilde{K}}_{\chi}^{2}\\
\mathcal{\tilde{K}}_{\pi}^{3} & \mathcal{\tilde{K}}_{\psi}^{3} & \mathcal{\tilde{K}}_{\phi}^{3} & 0\\
\mathcal{\tilde{K}}_{\pi}^{4} & \mathcal{\tilde{K}}_{\psi}^{4} & 0 & 0
\end{pmatrix}~,
\end{equation}
with the following coefficients
\begin{align}
    \mathcal{\tilde{W}}_{\ddot{\pi}}^{1}&=\frac{3}{8}M_{p}^{2}a^{3}d^{2}(t)~,
    &\mathcal{\tilde{W}}_{\ddot{\psi}}^{1}&=-\frac{3}{2}M_{p}^{2}a^{3}\Big(\epsilon c(t)+d(t)\Big)~,\\
    \mathcal{\tilde{W}}_{\ddot{\pi}}^{2}&=-\frac{3}{2}M_{p}^{2}a^{3}\Big(\epsilon c(t)+d(t)\Big)~,
    &\mathcal{\tilde{W}}_{\ddot{\psi}}^{2}&=6M_{p}^{2}a^{3}~,    
\end{align}
\begin{align}   
\mathcal{\tilde{K}}_{\pi}^{1}&= 
\frac{3}{8}M_{p}^{2}a^{3}d(t)\Big(3Hd(t)+2\dot{d} (t)\Big)\partial_{t} \nonumber 
\\ 
   &\ \ \ \ 
-\frac{3}{128}M_{p}^{2}a^{3}\Big\{36\epsilon^{3}c^{3}(t)H-3\epsilon^{2}c^{2}
(t)\Big[-3d^{2}(t)-36d(t)H\nonumber\\ \nonumber
   &\ \ \ \ 
+4\Big(-24H^{2}+\epsilon\dot{c}(t)+\dot{d}(t)\Big)\Big]+2d(t)\Big[9d^{3}
(t)+72Hd^{2}(t)\\  \nonumber
   &\ \ \ \ +12d(t)\Big(12H^{2}-\epsilon\dot{c}(t)\Big)-8\Big(3\epsilon H\dot{c}(t)+3H\dot{d}(t)+\epsilon\ddot{c}(t)+\ddot{d}\Big)\Big]\\ \nonumber
   &\ \ \ \ -2\epsilon 
c(t)\Big[-9d^{3}(t)-108d^{2}(t)H+d(t)\Big(-288H^{2}+12\epsilon\dot{c}(t)\Big)\\  
   &\ \ \ \  +8\Big(3\epsilon 
H\dot{c}(t)+3H\dot{d}(t)+\epsilon\ddot{c}+\ddot{d}\Big)\Big]\Big\}-\frac{3}{8}M_
{p}^{2}ad^{2}(t)\partial^{2}~,\\
   \mathcal{\tilde{K}}_{\psi}^{1}&=-\frac{9}{8}M_{p}^{2}a^{3}\Big(d^{2}(t)+8\epsilon c(t)H+8d(t)H\Big)\partial_{0}+M_{p}^{2}a\Big(\epsilon c(t)+d(t)\Big)\partial^{2}~,\\
   \mathcal{\tilde{K}}_{\phi}^{1}&=- 
\frac{3}{8}M_{p}^{2}a^{3}\Big(d^{2}(t)+4\epsilon c(t)H+4d(t)H\Big)\partial_{t} 
\nonumber\\ \nonumber
   &\ \ \ \ 
-\frac{3}{16}M_{p}^{2}a^{3}\Big[12\epsilon^{2}c^{2}(t)H+d(t)\Big(3d^{2}
(t)+24Hd(t)+48H^{2}-4\epsilon\dot{c}(t)\Big)\\   
   &\ \ \ \ +\epsilon c(t)\Big(3d^{2}(t)+24d(t)H+48H^{2}- 
4\epsilon\dot{c}(t)-4\dot{d}(t)\Big)\Big]-\frac{1}{2}M_{p}^{2}a\Big(\epsilon 
c(t)+d(t)\Big)\partial^{2}~,   \\
   \mathcal{\tilde{K}}_{\chi}^{1}&=-M_{p}^{2}a^{2}\Big(\epsilon 
c(t)+d(t)\Big)H\partial^{2}~,    
\end{align}
\begin{align}
   \mathcal{\tilde{K}}_{\pi}^{2}&= 
\frac{3}{8}M_{p}^{2}a^{3}\Big[3d^{2}(t)-8\Big(\epsilon\dot{c}(t)+\dot{d}
(t)\Big)\Big]\partial_{t} \nonumber\\ \nonumber
   &\ \ \ \ +\frac{8}{32}M_{p}^{2}a^{3}\Big[9d^{3}(t)+36\epsilon^{2}c^{2}(t)H+72d^{2}(t)H+12d(t)\Big(12H^{2}-\epsilon\dot{c}(t)+\dot{d}(t)\Big)\\ \nonumber
   &\ \ \ \ -3\epsilon 
c(t)\Big[-3d^{2}(t)-24d(t)H+4\Big(-12H^{2}+\epsilon\dot{c}(t)+\dot{d}
(t)\Big)\Big]\\  
   &\ \ \ \ -16\bigl(\epsilon\ddot{c}(t)+\ddot{d}(t)\Big)\Big\}+M_{p}^{2}a\Big(\epsilon c(t)+d(t)\Big)\partial^{2}~,\\
   \mathcal{\tilde{K}}_{\psi}^{2}&=18M_{p}^{2}a^{3}H\partial_{t}-2M_{p}^{2}a\partial^{2}~,\\
   \mathcal{\tilde{K}}_{\phi}^{2}&=
   \frac{3}{2}M_{p}^{2}a^{3}\Big(\epsilon c(t)+d(t)+4H\Big)\partial_{t} 
\nonumber\\  
   &\ \ \ \ +\frac{9}{8}M_{p}^{2}a^{3}\Big[d^{2}(t)+8d(t)H+8H\Big(\epsilon c(t)+2H\Big)\Big]+2M_{p}^{2}a\partial^{2}~,\\
   \mathcal{\tilde{K}}_{\chi}^{2}&=-M_{p}^{2}a^{2}\Big(\epsilon c(t)+d(t)\Big)\partial^{2}~,   
\end{align}
\begin{align}
   \mathcal{\tilde{K}}_{\pi}^{3}&= 
\frac{3}{8}M_{p}^{2}a^{3}\Big(d^{2}(t)+4Hd(t)+4\epsilon Hc(t)\Big)\partial_{t}  
\nonumber \\ \nonumber  
   &\ \ \ \ +\frac{1}{32}M_{p}^{2}a^{3}\Big\{-9d^{3}(t)\pi-36\epsilon^{2}c^{2}(t)H\pi-72d^{2}(t)H\pi+48H\pi\Big(\epsilon\dot{c}(t)+\dot{d}(t)\Big)\\ \nonumber  
   &\ \ \ \ +\epsilon 
c(t)\Big[-9d^{2}(t)\pi-72d(t)H\pi+12\pi\Big(-12H^{2}+\epsilon\dot{c}(t)+\dot{d}
(t)\Big)\Big]\Big\}\\
   &\ \ \ \ -12d(t)\Big(12H^{2} 
\pi-\epsilon\pi\dot{c}(t)-\pi\dot{d}(t)\Big)-\frac{1}{2}M_{p}^{2}a^{3}
\Big(\epsilon c(t)\partial^{2}+d(t)\Big)\partial^{2}~,\\  
   \mathcal{\tilde{K}}_{\psi}^{3}&=-\frac{3}{2}M_{p}^{2}a^{3}\Big(\epsilon c(t)+d(t)+4H\Big)\partial_{t}+2M_{p}^{2}a\partial^{2}~,\\
   \mathcal{\tilde{K}}_{\phi}^{3}&=-\frac{3}{8}M_{p}^{2}a^{3}\Big[d^{2}(t)+8d(t)H+8H\Big(\epsilon c(t)+2H\Big)\Big]~,  
\end{align}
and
\begin{align}
   \mathcal{\tilde{K}}_{\pi}^{4}&=-M_P^2 a^2\Big(\epsilon\, c(t)+d(t)\Big)H\partial^2~, &\mathcal{\tilde{K}}_{\psi}^{4}&=-M_P^2 a^2\Big(\epsilon\, c(t)+d(t)\Big)\partial^2~. 
\end{align}
As we mentioned in the main text, based on $\mathcal{\tilde E}_3^{(1)}=0$ and 
$\mathcal{\tilde E}_4^{(1)}=0$, we can solve $\phi$ and $\psi$ in terms of 
$\pi$. Then the Jacobian degenerates and induces a further constraint $\tilde 
C=0$, from which we can also resolve $\chi$ in terms of $\pi$ as
\begin{align}
\chi=\mathcal{\tilde C}_{\dot{\pi}}\dot{\pi}+\mathcal{\tilde C}_{\pi}\pi~,
\end{align}
where we have
\begin{align}
    \mathcal{C}_{\dot{\pi}}^{\text{M}}&=- \frac{20\epsilon\Big(2d(t)+\epsilon 
c(t)\Big)c(t)}{a\Big(d^{2}(t)+8Hd(t)+16H^{2}+8\epsilon 
Hc(t)\Big)\Big(d(t)+\epsilon c(t)\Big)}~,
\end{align}
and
\begin{align}
  \nonumber  \mathcal{C}_{\pi}^{\text{M}}&=\frac{8\epsilon\Big(d(t)+4H+\epsilon 
c(t) \Big)\Big(2d(t)+\epsilon 
c(t)\Big)c(t)}{3a^{3}\Big(d^{2}(t)+8Hd(t)+16H^{2}+8\epsilon 
Hc(t)\Big)^{2}\Big(d(t)+\epsilon c(t)\Big)}\partial^{2}\\ \nonumber
    &+\frac{1}{a\Big(d(t)+\epsilon c(t)\Big)\Big(d^{2}(t)+8Hd(t)+16H^{2}+8\epsilon Hc(t)+2H\Big)^{2}}\\ \nonumber
    &\ \ \ \ \times\bigg[d(t)\Big(d(t)+4H\Big)^{2}\Big(d^{2}(t)+8Hd(t)+16H^{2}-4\epsilon\dot{c}(t)\Big)\\ \nonumber
    &\ \ \ \ +\epsilon\Big(d(t)+4H\Big)\Big(-4\dot{d}(t)d(t)-16H\dot{d}(t)+7d^{3}(t)+48Hd^{2}(t)\\ \nonumber
    &\ \ \ \ \ \ \ \ \ \ \ \ +96H^{2}d(t)+64H^{3}-12\epsilon d(t)\dot{c}(t)-16\epsilon H\dot{c}(t)\Big)c(t)\\ \nonumber
    &\ \ \ \ +\epsilon^{2}\Big(-32H\dot{d}(t)-8\dot{d}(t)d(t)+9d^{3}(t)+86Hd^{2}(t)+288H^{2}d(t)\\ \nonumber
    &\ \ \ \ \ \ \ \ \ \ \ \ +224H^{3}-12\epsilon d(t)\dot{c}(t)-32\epsilon 
H\dot{c}(t)\Big)c^{2}(t)\\  
    &\ \ \ \  
+\epsilon^{3}\left(3d^{2}(t)+48Hd(t)+96H^{2}-4\epsilon\,\dot{c}(t)-4\dot{d}
(t)\right)c^{3}(t)+12\epsilon^{4}Hc^{4}(t)\bigg]~.
\end{align}

\section{Conformal transformation of the modified gravity model\label{appendixB}}

In this section, we consider the modified Lagrangian \eqref{model1} and apply the inverse conformal transformations to find the simpler form of the action in Jordon frame.  
We consider $d=-\frac{2}{\sqrt{3}}\dot\Psi_{0}$ and $c(t)=C(\Psi_{0})\dot\Psi_{0}$ 
where $C(\Psi)$ is a general function of $\Psi$.  
We obtain a generally covariant action from the modified coefficients as
\begin{equation}\label{covariant action with the modified coefficients}
    S=\frac{M_{P}^{2}}{2}\int\text{d}^{4}x\, 
\hat{e}\Big[-\hat{T}+(-\frac{2}{\sqrt{3}}+\epsilon 
C(\Psi))\hat{T}^{\mu}\nabla_{\mu}\Psi+\frac{1}{2}\hat{g}^{\mu\nu}\nabla_{\mu}
\Psi\nabla_{\nu}\Psi-V(\Psi)\Big]~,
\end{equation}
with
\begin{equation}
    V(\Psi)=[f^{\prime}(\tilde{\Psi}(\Psi))]^{-2}\Big[f(\tilde{\Psi}(\Psi))-\tilde{\Psi}(\Psi)f^{\prime}(\tilde{\Psi}(\Psi))\Big]~.
\end{equation}

As the first step, we shall follow an inverse procedure in Subsection \ref{Section:2.2} on the modified action.
In specific, a similar choice of the conformal factor in the $f(T)$ equivalent case will be taken. We recast $\Psi$ back to $\tilde\Psi$ using the relation 
$\Psi^\prime(\tilde{\Psi})=\sqrt{3}f^{\prime\prime}/f^{\prime}$, 
where we have used the conformal transformation and noted that 
$\nabla_{\mu}\Psi=\sqrt{3}f^{\prime\prime}(f^{\prime})^{-1}\nabla_{\mu}\tilde{
\Psi}$.
Using the behaviour of vector torsion and torsion scalar under conformal transformations
\begin{align}
    \hat{T}&=\Omega^{2}T+4\Omega g^{\mu\nu}\partial_{\nu}\Omega T_{\mu}-6g^{\mu\nu}\partial_{\mu}\Omega\partial_{\nu}\Omega~,\\
    \hat{T}^{\mu}&=\Omega^{2}T^{\mu}-3\Omega g^{\mu\rho}\partial_{\rho}\Omega~,
\end{align}
the action becomes
\begin{align}
         S&=\frac{M_{P}^{2}}{2}\int \text{d}^{4}x\, 
e\Big[-\Omega^{-2}T-4\Omega^{-3}g^{\mu\nu}\partial_{\nu}\Omega 
T_{\mu}+6\Omega^{-4}g^{\mu\nu}\partial_{\mu}\Omega\partial_{\nu}\Omega\nonumber 
\\ \nonumber 
        &\ \ \ \ +(-2+\sqrt{3}\epsilon 
C(\Psi))f^{\prime\prime}(f^{\prime})^{-1}(\Omega^{-2}T^{\mu}-3\Omega^{-3}g^{
\mu\rho}\partial_{\mu}\Omega)\nabla_{\mu}\tilde{\Psi}\\  
        &\ \ \ \ +\frac{3}{2}\Omega^{-4}(f^{\prime\prime})^{2}(f^{\prime})^{-2}\hat{g}^{\mu\nu}\nabla_{\mu}\tilde{\Psi}\nabla_{\nu}\tilde{\Psi}-\Omega^{-4}(f^{\prime})^{-2}(f-\tilde{\Psi}f^{\prime})\Big]~.
\end{align}
If we set the conformal factor as $\Omega=(f^{\prime}){}^{-1/2}$, we 
obtain the  form
\begin{equation}\label{effective action}
   S= \frac{M_{P}^{2}}{2}\int \text{d}^{4}x\, 
e\Big[-f^{\prime}T-V(\tilde{\Psi})-\epsilon\omega(\tilde{\Psi})g^{\mu\nu}\partial_{\mu}
\tilde{\Psi}\partial_{\nu}\tilde{\Psi}+\epsilon\tilde{c}(\tilde{\Psi})\partial_{
\mu}\tilde{\Psi}T^{\mu}\Big],
\end{equation}
with
\begin{align}
    &V(\tilde{\Psi})=f(\tilde{\Psi})-\tilde{\Psi}f^{\prime}(\tilde{\Psi})~,\\
    &\omega(\tilde{\Psi})=-\frac{3\sqrt{3}}{2}C(\tilde{\Psi})(f^{\prime\prime})^{2}(f^{\prime})^{-1}~,\\
    &\tilde{c}(\tilde{\Psi})=\sqrt{3}C(\tilde{\Psi})f^{\prime\prime}~,
\end{align}
as a modified form comparing to \eqref{fT Jordan frame}, namely
\begin{equation}
      S_{f(T)}=\frac{M_P^2}{2}\int  \text{d}^4x\, e \Big[-f^{\prime}T-V(\tilde{\Psi})\Big]~.
\end{equation}
Once we consider $\epsilon=0$, the action degenerates to the above action, 
which is equivalent to 
\begin{equation}
   S= \frac{M_{P}^{2}}{2}\int \text{d}^{4}x\, e\Big[\lambda (T-\tilde{\Psi})+f(\tilde{\Psi})\Big]~.
\end{equation}
We mention that since the action has been modified, the original relation 
$\lambda=f^\prime$ generated from variation with respect to $\tilde\Psi$ 
cannot be satisfied on-shell. 
Thus, we cannot recast the effective action \eqref{effective action} into original $f(T)$ gravity with some additional terms along this line.

In order to obtain a more compact form, in the following, we shall proceed by directly applying a general conformal transformation to the action \eqref{covariant action with the modified coefficients} without introducing the corresponding function $f$. As we find, the resulting action acquires the form
\begin{align}
    S&=\frac{M_{P}^{2}}{2}\int \text{d}^{4}x\,  
e\Big[-\Omega^{-2}T-4\Omega^{-3}T^{\mu}\nabla_{\mu}\Omega 
+\Big(-\frac{2}{\sqrt{3}}+\epsilon 
C(\Psi)\Big)\Omega^{-2}T^{\mu}\nabla_{\mu}\Psi\nonumber \\ 
    &\ \ \ \ +6\Omega^{-4}g^{\mu\nu}\nabla_{\mu}\Omega\nabla_{\nu} 
\Omega-(-2\sqrt{3}+3\epsilon 
C(\Psi))\Omega^{-3}g^{\mu\nu}\nabla_{\mu}\Omega\nabla_{\nu}\Psi+\frac{1}{2}
\Omega^{-2}g^{\mu\nu}\nabla_{\mu}\Psi\nabla_{\nu}\Psi-\Omega^{-4}V(\Psi)\Big]~.
\end{align}
Then, without loss of generality, we introduce a new auxiliary scalar field 
$\phi(x)$ to  replace the conformal factor. When we set the conformal factor 
as $\Omega=F^{-1/2}(\phi)$, where $F(\phi)$ is a general function of $\phi$, we 
have
\begin{align}
    S&=\frac{M_{P}^{2}}{2}\int \text{d}^{4}x\,  
e\Big[-FT+2F^{\prime}T^{\mu}\nabla_{\mu}\phi+(-\frac{2}{\sqrt{3}}+\epsilon 
C(\Psi))FT^{\mu}\nabla_{\mu}\Psi\nonumber\\ 
    &\ \ \ \ +\frac{3}{2}G^{-1}(F^{\prime}) 
^{2}g^{\mu\nu}\nabla_{\mu}\phi\nabla_{\nu}\phi+(-\sqrt{3}+\frac{3}{2}\epsilon 
C(\Psi))F^{\prime}g^{\mu\nu}\nabla_{\mu}\Psi\nabla_{\nu}\phi+\frac{1}{2}Fg^{
\mu\nu}\nabla_{\mu}\Psi\nabla_{\nu}\Psi-F^{2}V(\Psi)\Big]~.
\end{align}
We can always choose a conformal factor satisfying
\begin{equation}
    \frac{\text{d}\ln F}{\text{d}\phi}=\Big(-\frac{1}{2}\epsilon C(\Psi)+\frac{1}{\sqrt{3}}\Big)\frac{\text{d}\Psi}{\text{d}\phi}~,
\end{equation}
such that the coupling terms, namely $T^{\mu}\nabla_{\mu}\phi$ and 
$T^{\mu}\nabla_{\mu}\Phi$, vanish. Replacing $\nabla_{\mu}\Psi$ in the action 
with the relation
\begin{equation}
    \nabla_{\mu}\Psi=-\frac{2F^{\prime}}{(-\frac{2}{\sqrt{3}}+\epsilon C(\Psi))F}\nabla_{\mu}\phi~,
\end{equation}
yields
\begin{equation}
    S=\frac{M_{P}^{2}}{2}\int \text{d}^{4}x\, 
e\Big[-F(\phi)T-\omega(\phi)\nabla^{\mu}\phi\nabla_{\mu}\phi-F^{2}V(\phi)\Big]~,
\end{equation}
with
\begin{equation}
    \omega(\phi)=\frac{(F^{\prime})^{2}}{F}\Big[\frac{3}{2}-\frac{1}{(\epsilon C(\phi)-\frac{2}{\sqrt{3}})^{2}}\Big]~.
\end{equation}
Finally, we re-define the scalar field in order to make the kinetic term to 
obtain a canonical form through 
$\nabla_{\mu}\tilde{\phi}=2\omega^{1/2}\nabla_{\mu}\phi$, and we thus result to
\begin{equation}
    S=\frac{M_{P}^{2}}{2}\int \text{d}^{4}x\,  
e\Big[-F(\tilde\phi)T-\frac{1}{2}\nabla^{\mu}\tilde\phi\nabla_{\mu}
\tilde\phi-\tilde V(\tilde\phi)\Big]~,
\end{equation}
with
\begin{equation}
    \tilde V(\tilde\phi)=F^{2}(\tilde\phi)V(\tilde\phi)~.
\end{equation}

\bibliographystyle{Style}
\bibliography{ref}

\providecommand{\href}[2]{#2}\begingroup\raggedright\begin{thebibliography}{100}

\bibitem{CANTATA:2021ktz}
{\bf CANTATA} Collaboration, E.~Saridakis et~al., {\em {Modified Gravity and Cosmology: An Update by the CANTATA Network}}.
\newblock Springer, 2021.

\bibitem{Carroll:2003wy}
S.~M. Carroll, V.~Duvvuri, M.~Trodden, and M.~S. Turner, {\it {Is cosmic speed - up due to new gravitational physics?}},  Phys. Rev. D {\bf 70} (2004) 043528, [\href{http://arxiv.org/abs/astro-ph/0306438}{{\tt astro-ph/0306438}}].

\bibitem{Nojiri:2003ft}
S.~Nojiri and S.~D. Odintsov, {\it {Modified gravity with negative and positive powers of the curvature: Unification of the inflation and of the cosmic acceleration}},  Phys. Rev. D {\bf 68} (2003) 123512, [\href{http://arxiv.org/abs/hep-th/0307288}{{\tt hep-th/0307288}}].

\bibitem{Copeland:2006wr}
E.~J. Copeland, M.~Sami, and S.~Tsujikawa, {\it {Dynamics of dark energy}},  Int. J. Mod. Phys. D {\bf 15} (2006) 1753--1936, [\href{http://arxiv.org/abs/hep-th/0603057}{{\tt hep-th/0603057}}].

\bibitem{Aldrovandi:2013wha}
R.~Aldrovandi and J.~G. Pereira, {\em {Teleparallel Gravity: An Introduction}}.
\newblock Springer, Dordrechts, 2012.

\bibitem{Cai:2015emx}
Y.-F. Cai, S.~Capozziello, M.~De~Laurentis, and E.~N. Saridakis, {\it {f(T) teleparallel gravity and cosmology}},  Rept. Prog. Phys. {\bf 79} (2016), no.~10 106901, [\href{http://arxiv.org/abs/1511.07586}{{\tt arXiv:1511.07586}}].

\bibitem{Krssak:2018ywd}
M.~Kr\v{s}\v{s}\'ak, R.~J. van~den Hoogen, J.~G. Pereira, C.~G. B\"ohmer, and A.~A. Coley, {\it {Teleparallel theories of gravity: illuminating a fully invariant approach}},  Class. Quant. Grav. {\bf 36} (2019), no.~18 183001, [\href{http://arxiv.org/abs/1810.12932}{{\tt arXiv:1810.12932}}].

\bibitem{Bahamonde:2021gfp}
S.~Bahamonde, K.~F. Dialektopoulos, C.~Escamilla-Rivera, G.~Farrugia, V.~Gakis, M.~Hendry, M.~Hohmann, J.~Levi~Said, J.~Mifsud, and E.~Di~Valentino, {\it {Teleparallel gravity: from theory to cosmology}},  Rept. Prog. Phys. {\bf 86} (2023), no.~2 026901, [\href{http://arxiv.org/abs/2106.13793}{{\tt arXiv:2106.13793}}].

\bibitem{Ferraro:2006jd}
R.~Ferraro and F.~Fiorini, {\it {Modified teleparallel gravity: Inflation without inflaton}},  Phys. Rev. {\bf D75} (2007) 084031, [\href{http://arxiv.org/abs/gr-qc/0610067}{{\tt gr-qc/0610067}}].

\bibitem{Ferraro:2008ey}
R.~Ferraro and F.~Fiorini, {\it {On Born-Infeld Gravity in Weitzenbock spacetime}},  Phys. Rev. {\bf D78} (2008) 124019, [\href{http://arxiv.org/abs/0812.1981}{{\tt arXiv:0812.1981}}].

\bibitem{Bengochea:2008gz}
G.~R. Bengochea and R.~Ferraro, {\it {Dark torsion as the cosmic speed-up}},  Phys. Rev. {\bf D79} (2009) 124019, [\href{http://arxiv.org/abs/0812.1205}{{\tt arXiv:0812.1205}}].

\bibitem{Linder:2010py}
E.~V. Linder, {\it {Einstein's Other Gravity and the Acceleration of the Universe}},  Phys. Rev. {\bf D81} (2010) 127301, [\href{http://arxiv.org/abs/1005.3039}{{\tt arXiv:1005.3039}}]. [Erratum: Phys. Rev.D82,109902(2010)].

\bibitem{Maluf:2013gaa}
J.~W. Maluf, {\it {The teleparallel equivalent of general relativity}},  Annalen Phys. {\bf 525} (2013) 339--357, [\href{http://arxiv.org/abs/1303.3897}{{\tt arXiv:1303.3897}}].

\bibitem{Zheng:2010am}
R.~Zheng and Q.-G. Huang, {\it {Growth factor in $f(T)$ gravity}},  JCAP {\bf 03} (2011) 002, [\href{http://arxiv.org/abs/1010.3512}{{\tt arXiv:1010.3512}}].

\bibitem{Cai:2011tc}
Y.-F. Cai, S.-H. Chen, J.~B. Dent, S.~Dutta, and E.~N. Saridakis, {\it {Matter Bounce Cosmology with the f(T) Gravity}},  Class. Quant. Grav. {\bf 28} (2011) 215011, [\href{http://arxiv.org/abs/1104.4349}{{\tt arXiv:1104.4349}}].

\bibitem{Cardone:2012xq}
V.~F. Cardone, N.~Radicella, and S.~Camera, {\it {Accelerating f(T) gravity models constrained by recent cosmological data}},  Phys. Rev. D {\bf 85} (2012) 124007, [\href{http://arxiv.org/abs/1204.5294}{{\tt arXiv:1204.5294}}].

\bibitem{Krssak:2015oua}
M.~Kr\v{s}\v{s}\'ak and E.~N. Saridakis, {\it {The covariant formulation of f(T) gravity}},  Class. Quant. Grav. {\bf 33} (2016), no.~11 115009, [\href{http://arxiv.org/abs/1510.08432}{{\tt arXiv:1510.08432}}].

\bibitem{Chen:2019ftv}
Z.~Chen, W.~Luo, Y.-F. Cai, and E.~N. Saridakis, {\it {New test on general relativity and $f(T)$ torsional gravity from galaxy-galaxy weak lensing surveys}},  Phys. Rev. D {\bf 102} (2020), no.~10 104044, [\href{http://arxiv.org/abs/1907.12225}{{\tt arXiv:1907.12225}}].

\bibitem{Ren:2021uqb}
X.~Ren, Y.~Zhao, E.~N. Saridakis, and Y.-F. Cai, {\it {Deflection angle and lensing signature of covariant f(T) gravity}},  JCAP {\bf 10} (2021) 062, [\href{http://arxiv.org/abs/2105.04578}{{\tt arXiv:2105.04578}}].

\bibitem{Ren:2021tfi}
X.~Ren, T.~H.~T. Wong, Y.-F. Cai, and E.~N. Saridakis, {\it {Data-driven Reconstruction of the Late-time Cosmic Acceleration with f(T) Gravity}},  Phys. Dark Univ. {\bf 32} (2021) 100812, [\href{http://arxiv.org/abs/2103.01260}{{\tt arXiv:2103.01260}}].

\bibitem{dosSantos:2021owt}
F.~B.~M. dos Santos, J.~E. Gonzalez, and R.~Silva, {\it {Observational constraints on f(T) gravity from model-independent data}},  Eur. Phys. J. C {\bf 82} (2022), no.~9 823, [\href{http://arxiv.org/abs/2112.15249}{{\tt arXiv:2112.15249}}].

\bibitem{Zhao:2022gxl}
Y.~Zhao, X.~Ren, A.~Ilyas, E.~N. Saridakis, and Y.-F. Cai, {\it {Quasinormal modes of black holes in f(T) gravity}},  JCAP {\bf 10} (2022) 087, [\href{http://arxiv.org/abs/2204.11169}{{\tt arXiv:2204.11169}}].

\bibitem{Wang:2023qfm}
Q.~Wang, X.~Ren, B.~Wang, Y.-F. Cai, W.~Luo, and E.~N. Saridakis, {\it {Galaxy-galaxy lensing data: $f(T)$ gravity challenges General Relativity}},  \href{http://arxiv.org/abs/2312.17053}{{\tt arXiv:2312.17053}}.

\bibitem{Geng:2011aj}
C.-Q. Geng, C.-C. Lee, E.~N. Saridakis, and Y.-P. Wu, {\it {\textquotedblleft{}Teleparallel\textquotedblright{} dark energy}},  Phys. Lett. B {\bf 704} (2011) 384--387, [\href{http://arxiv.org/abs/1109.1092}{{\tt arXiv:1109.1092}}].

\bibitem{Geng:2011ka}
C.-Q. Geng, C.-C. Lee, and E.~N. Saridakis, {\it {Observational Constraints on Teleparallel Dark Energy}},  JCAP {\bf 01} (2012) 002, [\href{http://arxiv.org/abs/1110.0913}{{\tt arXiv:1110.0913}}].

\bibitem{Otalora:2014aoa}
G.~Otalora, {\it {A novel teleparallel dark energy model}},  Int. J. Mod. Phys. D {\bf 25} (2015), no.~02 1650025, [\href{http://arxiv.org/abs/1402.2256}{{\tt arXiv:1402.2256}}].

\bibitem{Bahamonde:2016kba}
S.~Bahamonde and C.~G. B\"ohmer, {\it {Modified teleparallel theories of gravity: Gauss\textendash{}Bonnet and trace extensions}},  Eur. Phys. J. C {\bf 76} (2016), no.~10 578, [\href{http://arxiv.org/abs/1606.05557}{{\tt arXiv:1606.05557}}].

\bibitem{Bahamonde:2017wwk}
S.~Bahamonde, C.~G. B\"ohmer, and M.~Kr\v{s}\v{s}\'ak, {\it {New classes of modified teleparallel gravity models}},  Phys. Lett. B {\bf 775} (2017) 37--43, [\href{http://arxiv.org/abs/1706.04920}{{\tt arXiv:1706.04920}}].

\bibitem{Hohmann:2017duq}
M.~Hohmann, L.~J\"arv, M.~Kr\v{s}\v{s}\'ak, and C.~Pfeifer, {\it {Teleparallel theories of gravity as analogue of nonlinear electrodynamics}},  Phys. Rev. D {\bf 97} (2018), no.~10 104042, [\href{http://arxiv.org/abs/1711.09930}{{\tt arXiv:1711.09930}}].

\bibitem{Hohmann:2018dqh}
M.~Hohmann and C.~Pfeifer, {\it {Scalar-torsion theories of gravity II: $L(T, X, Y, \phi)$ theory}},  Phys. Rev. D {\bf 98} (2018), no.~6 064003, [\href{http://arxiv.org/abs/1801.06536}{{\tt arXiv:1801.06536}}].

\bibitem{Hohmann:2018rwf}
M.~Hohmann, L.~J\"arv, and U.~Ualikhanova, {\it {Covariant formulation of scalar-torsion gravity}},  Phys. Rev. D {\bf 97} (2018), no.~10 104011, [\href{http://arxiv.org/abs/1801.05786}{{\tt arXiv:1801.05786}}].

\bibitem{Bahamonde:2022ohm}
S.~Bahamonde, K.~F. Dialektopoulos, M.~Hohmann, J.~Levi~Said, C.~Pfeifer, and E.~N. Saridakis, {\it {Perturbations in Non-Flat Cosmology for $f(T)$ gravity}},  \href{http://arxiv.org/abs/2203.00619}{{\tt arXiv:2203.00619}}.

\bibitem{Nester:1998mp}
J.~M. Nester and H.-J. Yo, {\it {Symmetric teleparallel general relativity}},  Chin. J. Phys. {\bf 37} (1999) 113, [\href{http://arxiv.org/abs/gr-qc/9809049}{{\tt gr-qc/9809049}}].

\bibitem{Adak:2005cd}
M.~Adak, M.~Kalay, and O.~Sert, {\it {Lagrange formulation of the symmetric teleparallel gravity}},  Int. J. Mod. Phys. D {\bf 15} (2006) 619--634, [\href{http://arxiv.org/abs/gr-qc/0505025}{{\tt gr-qc/0505025}}].

\bibitem{BeltranJimenez:2017tkd}
J.~Beltr\'an~Jim\'enez, L.~Heisenberg, and T.~Koivisto, {\it {Coincident General Relativity}},  Phys. Rev. D {\bf 98} (2018), no.~4 044048, [\href{http://arxiv.org/abs/1710.03116}{{\tt arXiv:1710.03116}}].

\bibitem{BeltranJimenez:2019tme}
J.~Beltr\'an~Jim\'enez, L.~Heisenberg, T.~S. Koivisto, and S.~Pekar, {\it {Cosmology in $f(Q)$ geometry}},  Phys. Rev. D {\bf 101} (2020), no.~10 103507, [\href{http://arxiv.org/abs/1906.10027}{{\tt arXiv:1906.10027}}].

\bibitem{Gakis:2019rdd}
V.~Gakis, M.~Kr\v{s}\v{s}\'ak, J.~Levi~Said, and E.~N. Saridakis, {\it {Conformal gravity and transformations in the symmetric teleparallel framework}},  Phys. Rev. D {\bf 101} (2020), no.~6 064024, [\href{http://arxiv.org/abs/1908.05741}{{\tt arXiv:1908.05741}}].

\bibitem{Lazkoz:2019sjl}
R.~Lazkoz, F.~S.~N. Lobo, M.~Ortiz-Ba\~nos, and V.~Salzano, {\it {Observational constraints of $f(Q)$ gravity}},  Phys. Rev. D {\bf 100} (2019), no.~10 104027, [\href{http://arxiv.org/abs/1907.13219}{{\tt arXiv:1907.13219}}].

\bibitem{Zhao:2021zab}
D.~Zhao, {\it {Covariant formulation of f(Q) theory}},  Eur. Phys. J. C {\bf 82} (2022), no.~4 303, [\href{http://arxiv.org/abs/2104.02483}{{\tt arXiv:2104.02483}}].

\bibitem{Heisenberg:2023lru}
L.~Heisenberg, {\it {Review on f(Q) gravity}},  Phys. Rept. {\bf 1066} (2024) 1--78, [\href{http://arxiv.org/abs/2309.15958}{{\tt arXiv:2309.15958}}].

\bibitem{Boehmer:2023knj}
C.~G. Boehmer, E.~Jensko, and R.~Lazkoz, {\it {Dynamical Systems Analysis of f(Q) Gravity}},  Universe {\bf 9} (2023), no.~4 166, [\href{http://arxiv.org/abs/2303.04463}{{\tt arXiv:2303.04463}}].

\bibitem{Li:2011rn}
M.~Li, R.-X. Miao, and Y.-G. Miao, {\it {Degrees of freedom of $f(T)$ gravity}},  JHEP {\bf 07} (2011) 108, [\href{http://arxiv.org/abs/1105.5934}{{\tt arXiv:1105.5934}}].

\bibitem{Ferraro:2018axk}
R.~Ferraro and M.~J. Guzm\'an, {\it {Quest for the extra degree of freedom in $f(T)$ gravity}},  Phys. Rev. D {\bf 98} (2018), no.~12 124037, [\href{http://arxiv.org/abs/1810.07171}{{\tt arXiv:1810.07171}}].

\bibitem{Blixt:2020ekl}
D.~Blixt, M.-J. Guzm\'an, M.~Hohmann, and C.~Pfeifer, {\it {Review of the Hamiltonian analysis in teleparallel gravity}},  Int. J. Geom. Meth. Mod. Phys. {\bf 18} (2021), no.~supp01 2130005, [\href{http://arxiv.org/abs/2012.09180}{{\tt arXiv:2012.09180}}].

\bibitem{Blagojevic:2020dyq}
M.~Blagojevi\'c and J.~M. Nester, {\it {Local symmetries and physical degrees of freedom in $f(T)$ gravity: a Dirac Hamiltonian constraint analysis}},  Phys. Rev. D {\bf 102} (2020), no.~6 064025, [\href{http://arxiv.org/abs/2006.15303}{{\tt arXiv:2006.15303}}].

\bibitem{Dent:2010nbw}
J.~B. Dent, S.~Dutta, and E.~N. Saridakis, {\it {f(T) gravity mimicking dynamical dark energy. Background and perturbation analysis}},  JCAP {\bf 01} (2011) 009, [\href{http://arxiv.org/abs/1010.2215}{{\tt arXiv:1010.2215}}].

\bibitem{Chen:2010va}
S.-H. Chen, J.~B. Dent, S.~Dutta, and E.~N. Saridakis, {\it {Cosmological perturbations in f(T) gravity}},  Phys. Rev. D {\bf 83} (2011) 023508, [\href{http://arxiv.org/abs/1008.1250}{{\tt arXiv:1008.1250}}].

\bibitem{Izumi:2012qj}
K.~Izumi and Y.~C. Ong, {\it {Cosmological Perturbation in f(T) Gravity Revisited}},  JCAP {\bf 06} (2013) 029, [\href{http://arxiv.org/abs/1212.5774}{{\tt arXiv:1212.5774}}].

\bibitem{Chen:2014qtl}
P.~Chen, K.~Izumi, J.~M. Nester, and Y.~C. Ong, {\it {Remnant Symmetry, Propagation and Evolution in $f$(T) Gravity}},  Phys. Rev. D {\bf 91} (2015), no.~6 064003, [\href{http://arxiv.org/abs/1412.8383}{{\tt arXiv:1412.8383}}].

\bibitem{BeltranJimenez:2020lee}
J.~Beltr\'an~Jim\'enez and A.~Jim\'enez-Cano, {\it {On the strong coupling of Einsteinian Cubic Gravity and its generalisations}},  JCAP {\bf 01} (2021) 069, [\href{http://arxiv.org/abs/2009.08197}{{\tt arXiv:2009.08197}}].

\bibitem{BeltranJimenez:2021auj}
J.~Beltr\'an~Jim\'enez and T.~S. Koivisto, {\it {Accidental gauge symmetries of Minkowski spacetime in Teleparallel theories}},  Universe {\bf 7} (2021), no.~5 143, [\href{http://arxiv.org/abs/2104.05566}{{\tt arXiv:2104.05566}}].

\bibitem{Shabani:2023xfn}
H.~Shabani, A.~De, T.-H. Loo, and E.~N. Saridakis, {\it {Cosmology of f(Q) gravity in non-flat Universe}},  Eur. Phys. J. C {\bf 84} (2024), no.~3 285, [\href{http://arxiv.org/abs/2306.13324}{{\tt arXiv:2306.13324}}].

\bibitem{Gomes:2023tur}
D.~A. Gomes, J.~Beltr\'an~Jim\'enez, A.~J. Cano, and T.~S. Koivisto, {\it {Pathological Character of Modifications to Coincident General Relativity: Cosmological Strong Coupling and Ghosts in f(Q) Theories}},  Phys. Rev. Lett. {\bf 132} (2024), no.~14 141401, [\href{http://arxiv.org/abs/2311.04201}{{\tt arXiv:2311.04201}}].

\bibitem{Golovnev:2018wbh}
A.~Golovnev and T.~Koivisto, {\it {Cosmological perturbations in modified teleparallel gravity models}},  JCAP {\bf 11} (2018) 012, [\href{http://arxiv.org/abs/1808.05565}{{\tt arXiv:1808.05565}}].

\bibitem{Hohmann:2020vcv}
M.~Hohmann, {\it {General cosmological perturbations in teleparallel gravity}},  Eur. Phys. J. Plus {\bf 136} (2021), no.~1 65, [\href{http://arxiv.org/abs/2011.02491}{{\tt arXiv:2011.02491}}].

\bibitem{Golovnev:2020aon}
A.~Golovnev, {\it {Perturbations in $f(\mathbb{T})$ cosmology and the spin connection}},  JCAP {\bf 04} (2020) 014, [\href{http://arxiv.org/abs/2001.10015}{{\tt arXiv:2001.10015}}].

\bibitem{Deffayet:2005ys}
C.~Deffayet and J.-W. Rombouts, {\it {Ghosts, strong coupling and accidental symmetries in massive gravity}},  Phys. Rev. D {\bf 72} (2005) 044003, [\href{http://arxiv.org/abs/gr-qc/0505134}{{\tt gr-qc/0505134}}].

\bibitem{Blas:2009yd}
D.~Blas, O.~Pujolas, and S.~Sibiryakov, {\it {On the Extra Mode and Inconsistency of Horava Gravity}},  JHEP {\bf 10} (2009) 029, [\href{http://arxiv.org/abs/0906.3046}{{\tt arXiv:0906.3046}}].

\bibitem{Cai:2009dx}
R.-G. Cai, B.~Hu, and H.-B. Zhang, {\it {Dynamical Scalar Degree of Freedom in Horava-Lifshitz Gravity}},  Phys. Rev. D {\bf 80} (2009) 041501, [\href{http://arxiv.org/abs/0905.0255}{{\tt arXiv:0905.0255}}].

\bibitem{Hu:2022anq}
K.~Hu, T.~Katsuragawa, and T.~Qiu, {\it {ADM formulation and Hamiltonian analysis of f(Q) gravity}},  Phys. Rev. D {\bf 106} (2022), no.~4 044025, [\href{http://arxiv.org/abs/2204.12826}{{\tt arXiv:2204.12826}}].

\bibitem{Tomonari:2023wcs}
K.~Tomonari and S.~Bahamonde, {\it {Dirac\textendash{}Bergmann analysis and degrees of freedom of coincident f(Q)-gravity}},  Eur. Phys. J. C {\bf 84} (2024), no.~4 349, [\href{http://arxiv.org/abs/2308.06469}{{\tt arXiv:2308.06469}}].

\bibitem{Hu:2023gui}
K.~Hu, M.~Yamakoshi, T.~Katsuragawa, S.~Nojiri, and T.~Qiu, {\it {Nonpropagating ghost in covariant f(Q) gravity}},  Phys. Rev. D {\bf 108} (2023), no.~12 124030, [\href{http://arxiv.org/abs/2310.15507}{{\tt arXiv:2310.15507}}].

\bibitem{Zhao:2024kri}
D.~Zhao, {\it {Conformal transformation of f(Q) gravity and its cosmological perturbations}},  \href{http://arxiv.org/abs/2404.16299}{{\tt arXiv:2404.16299}}.

\bibitem{Li:2018ixg}
C.~Li, Y.~Cai, Y.-F. Cai, and E.~N. Saridakis, {\it {The effective field theory approach of teleparallel gravity, $f(T)$ gravity and beyond}},  JCAP {\bf 10} (2018) 001, [\href{http://arxiv.org/abs/1803.09818}{{\tt arXiv:1803.09818}}].

\bibitem{Cai:2018rzd}
Y.-F. Cai, C.~Li, E.~N. Saridakis, and L.~Xue, {\it {$f(T)$ gravity after GW170817 and GRB170817A}},  Phys. Rev. D {\bf 97} (2018), no.~10 103513, [\href{http://arxiv.org/abs/1801.05827}{{\tt arXiv:1801.05827}}].

\bibitem{Cai:2019bdh}
Y.-F. Cai, M.~Khurshudyan, and E.~N. Saridakis, {\it {Model-independent reconstruction of $f(T)$ gravity from Gaussian Processes}},  Astrophys. J. {\bf 888} (2020) 62, [\href{http://arxiv.org/abs/1907.10813}{{\tt arXiv:1907.10813}}].

\bibitem{Yan:2019gbw}
S.-F. Yan, P.~Zhang, J.-W. Chen, X.-Z. Zhang, Y.-F. Cai, and E.~N. Saridakis, {\it {Interpreting cosmological tensions from the effective field theory of torsional gravity}},  Phys. Rev. D {\bf 101} (2020), no.~12 121301, [\href{http://arxiv.org/abs/1909.06388}{{\tt arXiv:1909.06388}}].

\bibitem{Ren:2022aeo}
X.~Ren, S.-F. Yan, Y.~Zhao, Y.-F. Cai, and E.~N. Saridakis, {\it {Gaussian processes and effective field theory of $f(T)$ gravity under the $H_0$ tension}},  Astrophys. J. {\bf 932} (3, 2022) 131, [\href{http://arxiv.org/abs/2203.01926}{{\tt arXiv:2203.01926}}].

\bibitem{Mylova:2022ljr}
M.~Mylova, J.~Levi~Said, and E.~N. Saridakis, {\it {General Effective Field Theory of Teleparallel Gravity}},  \href{http://arxiv.org/abs/2211.11420}{{\tt arXiv:2211.11420}}.

\bibitem{Hu:2023xcf}
Y.-M. Hu, Y.~Yu, Y.-F. Cai, and X.~Gao, {\it {The effective field theory approach to the strong coupling issue in $f(T)$ gravity with a non-minimally coupled scalar field}},  \href{http://arxiv.org/abs/2311.12645}{{\tt arXiv:2311.12645}}.

\bibitem{Yang:2024kdo}
Y.~Yang, X.~Ren, Q.~Wang, Z.~Lu, D.~Zhang, Y.-F. Cai, and E.~N. Saridakis, {\it {Quintom cosmology and modified gravity after DESI 2024}},  Sci. Bull. (4, 2024) [\href{http://arxiv.org/abs/2404.19437}{{\tt arXiv:2404.19437}}].

\bibitem{Hu:2023juh}
Y.-M. Hu, Y.~Zhao, X.~Ren, B.~Wang, E.~N. Saridakis, and Y.-F. Cai, {\it {The effective field theory approach to the strong coupling issue in f(T) gravity}},  JCAP {\bf 07} (2023) 060, [\href{http://arxiv.org/abs/2302.03545}{{\tt arXiv:2302.03545}}].

\bibitem{dicke1962mach}
R.~H. Dicke, {\it Mach's principle and invariance under transformation of units},  Physical review {\bf 125} (1962), no.~6 2163.

\bibitem{Maeda:1988ab}
K.-i. Maeda, {\it {Towards the Einstein-Hilbert Action via Conformal Transformation}},  Phys. Rev. D {\bf 39} (1989) 3159.

\bibitem{Deruelle:2010ht}
N.~Deruelle and M.~Sasaki, {\it {Conformal equivalence in classical gravity: the example of 'Veiled' General Relativity}},  Springer Proc. Phys. {\bf 137} (2011) 247--260, [\href{http://arxiv.org/abs/1007.3563}{{\tt arXiv:1007.3563}}].

\bibitem{Paliathanasis:2023gfq}
A.~Paliathanasis, {\it {The Brans\textendash{}Dicke field in non-metricity gravity: cosmological solutions and conformal transformations}},  Eur. Phys. J. C {\bf 84} (2024), no.~2 125, [\href{http://arxiv.org/abs/2310.16357}{{\tt arXiv:2310.16357}}].

\bibitem{Yang:2010ji}
R.-J. Yang, {\it {Conformal transformation in $f(T)$ theories}},  EPL {\bf 93} (2011), no.~6 60001, [\href{http://arxiv.org/abs/1010.1376}{{\tt arXiv:1010.1376}}].

\bibitem{Hohmann:2018vle}
M.~Hohmann, {\it {Scalar-torsion theories of gravity I: general formalism and conformal transformations}},  Phys. Rev. D {\bf 98} (2018), no.~6 064002, [\href{http://arxiv.org/abs/1801.06528}{{\tt arXiv:1801.06528}}].

\bibitem{Bamba:2013jqa}
K.~Bamba, S.~D. Odintsov, and D.~S\'aez-G\'omez, {\it {Conformal symmetry and accelerating cosmology in teleparallel gravity}},  Phys. Rev. D {\bf 88} (2013) 084042, [\href{http://arxiv.org/abs/1308.5789}{{\tt arXiv:1308.5789}}].

\bibitem{Jirousek:2022jhh}
P.~Jirou\v{s}ek, K.~Shimada, A.~Vikman, and M.~Yamaguchi, {\it {New dynamical degrees of freedom from invertible transformations}},  JHEP {\bf 07} (2023) 154, [\href{http://arxiv.org/abs/2208.05951}{{\tt arXiv:2208.05951}}].

\bibitem{Flanagan:2004bz}
E.~E. Flanagan, {\it {The Conformal frame freedom in theories of gravitation}},  Class. Quant. Grav. {\bf 21} (2004) 3817, [\href{http://arxiv.org/abs/gr-qc/0403063}{{\tt gr-qc/0403063}}].

\bibitem{Faraoni:2006fx}
V.~Faraoni and S.~Nadeau, {\it {The (pseudo)issue of the conformal frame revisited}},  Phys. Rev. D {\bf 75} (2007) 023501, [\href{http://arxiv.org/abs/gr-qc/0612075}{{\tt gr-qc/0612075}}].

\bibitem{Chiba:2013mha}
T.~Chiba and M.~Yamaguchi, {\it {Conformal-Frame (In)dependence of Cosmological Observations in Scalar-Tensor Theory}},  JCAP {\bf 10} (2013) 040, [\href{http://arxiv.org/abs/1308.1142}{{\tt arXiv:1308.1142}}].

\bibitem{Kamenshchik:2014waa}
A.~Y. Kamenshchik and C.~F. Steinwachs, {\it {Question of quantum equivalence between Jordan frame and Einstein frame}},  Phys. Rev. D {\bf 91} (2015), no.~8 084033, [\href{http://arxiv.org/abs/1408.5769}{{\tt arXiv:1408.5769}}].

\bibitem{Faraoni:1998qx}
V.~Faraoni, E.~Gunzig, and P.~Nardone, {\it {Conformal transformations in classical gravitational theories and in cosmology}},  Fund. Cosmic Phys. {\bf 20} (1999) 121, [\href{http://arxiv.org/abs/gr-qc/9811047}{{\tt gr-qc/9811047}}].

\bibitem{DeFelice:2010aj}
A.~De~Felice and S.~Tsujikawa, {\it {f(R) theories}},  Living Rev. Rel. {\bf 13} (2010) 3, [\href{http://arxiv.org/abs/1002.4928}{{\tt arXiv:1002.4928}}].

\bibitem{Wright:2016ayu}
M.~Wright, {\it {Conformal transformations in modified teleparallel theories of gravity revisited}},  Phys. Rev. D {\bf 93} (2016), no.~10 103002, [\href{http://arxiv.org/abs/1602.05764}{{\tt arXiv:1602.05764}}].

\bibitem{Jarv:2018bgs}
L.~J\"arv, M.~R\"unkla, M.~Saal, and O.~Vilson, {\it {Nonmetricity formulation of general relativity and its scalar-tensor extension}},  Phys. Rev. D {\bf 97} (2018), no.~12 124025, [\href{http://arxiv.org/abs/1802.00492}{{\tt arXiv:1802.00492}}].

\bibitem{Zumalacarregui:2013pma}
M.~Zumalac\'arregui and J.~Garc\'\i{}a-Bellido, {\it {Transforming gravity: from derivative couplings to matter to second-order scalar-tensor theories beyond the Horndeski Lagrangian}},  Phys. Rev. D {\bf 89} (2014) 064046, [\href{http://arxiv.org/abs/1308.4685}{{\tt arXiv:1308.4685}}].

\bibitem{Domenech:2015tca}
G.~Dom\`enech, S.~Mukohyama, R.~Namba, A.~Naruko, R.~Saitou, and Y.~Watanabe, {\it {Derivative-dependent metric transformation and physical degrees of freedom}},  Phys. Rev. D {\bf 92} (2015), no.~8 084027, [\href{http://arxiv.org/abs/1507.05390}{{\tt arXiv:1507.05390}}].

\bibitem{Takahashi:2017zgr}
K.~Takahashi, H.~Motohashi, T.~Suyama, and T.~Kobayashi, {\it {General invertible transformation and physical degrees of freedom}},  Phys. Rev. D {\bf 95} (2017), no.~8 084053, [\href{http://arxiv.org/abs/1702.01849}{{\tt arXiv:1702.01849}}].

\bibitem{Krssak:2024xeh}
M.~Kr\v{s}\v{s}\'ak, {\it {Teleparallel Gravity, Covariance and Their Geometrical Meaning}},  in {\em {Tribute to Ruben Aldrovandi}} ({F. Caruso, J.G. Pereira and A. Santoro}, ed.).
\newblock {Editora Livraria da Física}, São Paulo, 2024.
\newblock \href{http://arxiv.org/abs/2401.08106}{{\tt arXiv:2401.08106}}.

\bibitem{Hehl:1994ue}
F.~W. Hehl, J.~D. McCrea, E.~W. Mielke, and Y.~Ne'eman, {\it {Metric affine gauge theory of gravity: Field equations, Noether identities, world spinors, and breaking of dilation invariance}},  Phys. Rept. {\bf 258} (1995) 1--171, [\href{http://arxiv.org/abs/gr-qc/9402012}{{\tt gr-qc/9402012}}].

\bibitem{Obukhov:2002tm}
Y.~N. Obukhov and J.~G. Pereira, {\it {Metric affine approach to teleparallel gravity}},  Phys. Rev. D {\bf 67} (2003) 044016, [\href{http://arxiv.org/abs/gr-qc/0212080}{{\tt gr-qc/0212080}}].

\bibitem{Ferraro:2011us}
R.~Ferraro and F.~Fiorini, {\it {Non trivial frames for f(T) theories of gravity and beyond}},  Phys. Lett. {\bf B702} (2011) 75--80, [\href{http://arxiv.org/abs/1103.0824}{{\tt arXiv:1103.0824}}].

\bibitem{Tamanini:2012hg}
N.~Tamanini and C.~G. Boehmer, {\it {Good and bad tetrads in f(T) gravity}},  Phys. Rev. D {\bf 86} (2012) 044009, [\href{http://arxiv.org/abs/1204.4593}{{\tt arXiv:1204.4593}}].

\bibitem{BeltranJimenez:2020fvy}
J.~Beltr\'an~Jim\'enez, A.~Golovnev, T.~Koivisto, and H.~Veerm\"ae, {\it {Minkowski space in $f(T)$ gravity}},  Phys. Rev. D {\bf 103} (2021), no.~2 024054, [\href{http://arxiv.org/abs/2004.07536}{{\tt arXiv:2004.07536}}].

\bibitem{Li:2010cg}
B.~Li, T.~P. Sotiriou, and J.~D. Barrow, {\it {$f(T)$ gravity and local Lorentz invariance}},  Phys. Rev. D {\bf 83} (2011) 064035, [\href{http://arxiv.org/abs/1010.1041}{{\tt arXiv:1010.1041}}].

\bibitem{Maluf:2011kf}
J.~W. Maluf and F.~F. Faria, {\it {Conformally invariant teleparallel theories of gravity}},  Phys. Rev. D {\bf 85} (2012) 027502, [\href{http://arxiv.org/abs/1110.3095}{{\tt arXiv:1110.3095}}].

\bibitem{Wei:2011yr}
H.~Wei, {\it {Dynamics of Teleparallel Dark Energy}},  Phys. Lett. B {\bf 712} (2012) 430--436, [\href{http://arxiv.org/abs/1109.6107}{{\tt arXiv:1109.6107}}].

\bibitem{Otalora:2013tba}
G.~Otalora, {\it {Scaling attractors in interacting teleparallel dark energy}},  JCAP {\bf 07} (2013) 044, [\href{http://arxiv.org/abs/1305.0474}{{\tt arXiv:1305.0474}}].

\bibitem{Dil:2015eum}
E.~Dil and E.~Kolay, {\it {Dynamics of Mixed Dark Energy Domination in Teleparallel Gravity and Phase-Space Analysis}},  Adv. High Energy Phys. {\bf 2015} (2015) 608252, [\href{http://arxiv.org/abs/1610.04608}{{\tt arXiv:1610.04608}}].

\bibitem{Bahamonde:2018miw}
S.~Bahamonde, M.~Marciu, and P.~Rudra, {\it {Generalised teleparallel quintom dark energy non-minimally coupled with the scalar torsion and a boundary term}},  JCAP {\bf 04} (2018) 056, [\href{http://arxiv.org/abs/1802.09155}{{\tt arXiv:1802.09155}}].

\end{thebibliography}\endgroup

\end{document}